\documentclass[12pt]{article} 
\pdfoutput=1
\usepackage[hyperfootnotes=false]{hyperref}
 \usepackage{amsmath}
\usepackage{amssymb}
\usepackage{graphicx}
\usepackage{tikz}
\usepackage{xcolor}
\usepackage{framed}
  \usepackage{subfigure}
  \usepackage{caption}
 \usepackage{mciteplus}
 \usepackage{bbold}
  
  \newlength{\abstractwidth}
  \newcommand{\be}{\begin{equation}}
  \newcommand{\bea}{\begin{eqnarray}}
  \newcommand{\eea}{\end{eqnarray}}
  \newcommand{\beq}{\begin{equation}}
  \newcommand{\ee}{\end{equation}}
  \newcommand{\eeq}{\end{equation}}

\DeclareMathOperator{\Tr}{Tr}
\newcommand{\id}{\mathbb{1}}
\newcommand{\ZZ}{\mathbb{Z}}
\newcommand{\RR}{\mathbb{R}}
\newcommand{\op}{\mathcal{O}}
\renewcommand{\Re}{\operatorname{Re}}
\renewcommand{\Im}{\operatorname{Im}}
\newcommand{\lb}{\langle}
\newcommand{\rb}{\rangle}

\newcommand{\modS}{\mathbb{S}}		
\newcommand{\fusion}{\mathbb{F}}	
\newcommand{\braid}{\mathbb{R}}		
\newcommand{\sbmatrix}[1]{
{\tiny\arraycolsep=0.3\arraycolsep\ensuremath{\begin{bmatrix}#1\end{bmatrix}}}
}	
\newcommand{\sixj}[6]{\Bigl\{\mbox{\small$\!\begin{array}{ccc} #1 \! & \!\! #3 \! & \!\! #5 \hspace{-2pt} \\[-1mm]  #2 \!  & \!\! #4 \!  & \!\! #6 \hspace{-2pt} \end{array}\!$}\!\Bigr\}}

\begin{document}

\begin{titlepage}
  \bigskip

  \bigskip\bigskip

  \bigskip

\begin{center}
{\Large \bf {A universal Schwarzian sector in two-dimensional }}
 \bigskip 
 ~~\\
{\Large \bf {conformal field theories }} 
    \bigskip
\bigskip
\end{center}

  \begin{center}

 \bf {Animik Ghosh$^{1,2}$, Henry Maxfield$^3$ and Gustavo J. Turiaci$^3$   }
  \bigskip \rm
\bigskip

{\small ${}^1$ Department of Physics and Astronomy, University of Kentucky, Lexington, KY 40506, USA}  \\
\vspace{0.2cm}
{\small  \hspace{-0.43cm} ${}^2$ Kavli Institute for Theoretical Physics, University of California, Santa Barbara, CA 93106, USA}  \\
\vspace{0.2cm}
{\small ${}^3$ Physics Department, University of California, Santa Barbara, CA 93106, USA}  \\
\rm

  \bigskip \rm
\bigskip
 
 \texttt{animik.ghosh@uky.edu, hmaxfield@ucsb.edu, turiaci@ucsb.edu }
\rm

\bigskip
\bigskip

  \end{center}

\vspace{2.5cm}
  \begin{abstract}

We show that an extremely generic class of two-dimensional conformal field theories (CFTs) contains a sector described by the Schwarzian theory. This applies to theories with no additional symmetries and large central charge, but does not require a holographic dual. Specifically, we use bootstrap methods to show that in the grand canonical ensemble, at low temperature with a chemical potential sourcing large angular momentum, the density of states and correlation functions are determined by the Schwarzian theory, up to parametrically small corrections. In particular, we compute out-of-time-order correlators in a controlled approximation. For holographic theories, these results have a gravitational interpretation in terms of large, near-extremal rotating BTZ black holes, which have a near horizon throat with nearly AdS$_2 \times S^1$ geometry. The Schwarzian describes strongly coupled gravitational dynamics in the throat, which can be reduced to Jackiw-Teitelboim (JT) gravity interacting with a $U(1)$ field associated to transverse rotations, coupled to matter. We match the physics in the throat to observables at the AdS$_3$ boundary, reproducing the CFT results.

 \medskip
  \noindent
  \end{abstract}
\bigskip \bigskip \bigskip

  \end{titlepage}

   \tableofcontents


\newpage
\section{Introduction }

Our understanding of near-extremal black holes has been recently revolutionized by the improved understanding of a universal dynamics which emerges at low temperature \cite{Almheiri:2014cka, Sachdev:2015efa, Jensen:2016pah, Maldacena:2016upp, Engelsoy:2016xyb, Cvetic:2016eiv}. It has long been known that black holes near extremality develop a long AdS$_2$ throat near the event horizon, which behaves rather differently from the analogous higher-dimensional AdS regions near black branes \cite{Preskill:1991tb,Maldacena:1998uz}. The underlying reason for this is that the AdS$_2$ region does not completely decouple from the physics far from the black hole in the low-temperature limit. Instead, there is a single mode which becomes increasingly important at low temperature, specifying the relationship between the time in AdS$_2$ and the time far from the black hole. This mode is governed by the Schwarzian theory (described gravitationally inside the AdS$_2$ throat by Jackiw-Teitelboim gravity \cite{Jackiw:1984je, Teitelboim:1983ux}), with action
\begin{equation}
	I_\text{Schw} = -C \int_0^\beta dt_E \, \left\{ \tan\left( \tfrac{\pi}{\beta} f(t_E)\right),t_E\right\},
\end{equation}
where $t_E$ is the asymptotic (Euclidean) time and $f(t_E)$ a time measured in the AdS$_2$ region, with $\{\cdot,\cdot\}$ denoting the Schwarzian derivative. The coefficient $C$, with dimensions of time, marks the inverse temperature $\beta$ at which the Schwarzian dynamics becomes strongly coupled, with quantum fluctuations of $f$ unsuppressed. This is a theory of a pseudo-Goldstone mode, determined by the nature of the symmetry breaking, which therefore appears in more general circumstances, not least the SYK model \cite{Sachdev:1992fk, Kitaev:2017awl, Polchinski:2016xgd, Maldacena:2016hyu, Jevicki:2016ito, Jevicki:2016bwu}.

In this paper, we show that, in extremely generic circumstances, two-dimensional conformal field theories (CFTs) contain a sector described by the Schwarzian theory. This sector has a gravitational description in terms of a near-horizon AdS$_2$ region of near-extremal rotating BTZ black holes, but exists even in theories without a local weakly coupled AdS$_3$ dual!

Since our results are very general, we will not make use of details of a particular theory to derive the Schwarzian.\footnote{The work in this paper is orthogonal to previous constructions of SYK-like models in two dimensions \cite{Berkooz:2016cvq, Turiaci:2017zwd, Murugan:2017eto}} Rather, we will use conformal bootstrap methods to study observables in states that enhance the effects of the Schwarzian sector, with very large angular momentum and low temperature. More specifically, we explicitly construct correlation functions in an appropriate limit of the grand canonical ensemble for angular momentum, requiring only a theory with a large central charge $c\gg 1$ and no conserved currents besides those of local conformal symmetry. With these general assumptions, we will show that the density of states and all correlators are dictated by the Schwarzian theory, with parametrically small corrections. To achieve this, we rigorously demonstrate that the correlators are dominated by a Virasoro identity block in an appropriate channel, before showing that the block reduces to the Schwarzian correlator in the appropriate limit. The methods for the latter calculation are much the same as used in \cite{Mertens:2017mtv} to compute correlation functions of the Schwarzian at strong coupling from Liouville theory, though the interpretation in this paper is rather different. This result is a striking example of the universality of gravity as a description of chaotic quantum systems.

The paper is organized as follows. For the remainder of the introduction, we will summarise our main results. In section \ref{sec:inv}, we discuss the partition function and spectrum of BTZ black holes and of CFTs in the limit of interest, which illustrates the main ideas while avoiding too many technical details. The main results appear in section \ref{sec:CFT}, where we show how correlation functions of CFTs reduce to those of the Schwarzian. Finally, in section \ref{sec:BTZ} we study near-extremal rotating BTZ black holes, to obtain a detailed gravitational interpretation of our CFT calculations.

\subsection{A near-extremal limit of CFT$_2$}

For our main results, we study correlation functions of two-dimensional, compact unitary CFTs with large central charge $c\gg 1$ in the grand canonical ensemble. Instead of using a temperature and chemical potential for angular momentum $J$, this ensemble can be described by independent temperatures $T_L$, $T_R$ for left- and right-moving conformal dimensions $h,\bar{h}$. We study a regime which we call the near-extremal limit, where $T_L$ is of order $c^{-1}$, and $T_R$ is taken to be very large. This ensures that the physics is dominated by states with very large angular momentum $J\gg c$ (controlled by the large $T_R$) and low temperature $T$ of order $c^{-1}$ (controlled by the small $T_L$). These states are related to CFT operators with a very large right-moving scaling dimension $\bar{h} \sim J$, and left-moving dimension close to $\frac{c-1}{24}$, with $h- \frac{c-1}{24}$ of order $c^{-1}$.

The physics in this regime is exemplified by the simplest correlator, namely a two-point function of light operators in the grand canonical ensemble. In our near-extremal limit, we find that the real-time two-point function (using light-cone coordinates $z=\varphi-t$ and $\bar{z}=\varphi+t$)  is given by
\begin{equation}\label{eqCFTSch}
	G^{\text{(CFT)}}(z,\bar{z};\beta_L,\beta_R) \sim \left(\frac{c}{6}\right)^{-2h} \left[\frac{1}{\pi T_R}\sinh\left(\pi T_R\bar{z}\right)\right]^{-2\bar{h}} G^{(\text{Schw})}_h(-z),
\end{equation}
where $G^{(\text{Schw})}_h(-z)$ is the exact (that is, strong coupling) real-time Schwarzian two point function at a temperature proportional to $T_L$. This is valid for large time separations $t$ of order the inverse temperature $T_L^{-1}$ (and for $\bar{z}$ of order one, but not too large, as described in section \eqref{sec:rmsdvb}). Note that the Schwarzian appears most naturally in real time, not Euclidean time.\footnote{There is another regime where the Euclidean time Schwarzian appears, related by a modular transformation. Taking very low or zero right-moving temperature, and left-moving temperature of order $c$, the left-moving block gives a Schwarzian evaluated at Euclidean time $z$.}

To show this, we first use the limit of large $T_R$ and modular invariance: in a modular S-dual channel, this becomes very small right-moving temperature, which effectively projects the sum over intermediate states onto the smallest value of $\bar{h}$, corresponding to the vacuum state and its left-moving descendants. This is where the restrictions on our theory are important, namely that it is unitary (so all operators have $h,\bar{h}\geq 0$), compact (having a unique $\mathfrak{sl}(2)$-invariant vacuum state, with $h=\bar{h}=0$),  and has a twist gap (so no primary states besides the vacuum have $\bar{h}=0$). With mild kinematic restrictions on $\bar{z}$, we conclude that the correlation function is given in this limit by a single conformal block with the vacuum state exchanged. This explains why the result \eqref{eqCFTSch} has factorised dependence on left- and right-moving variables $z,T_L$ and $\bar{z},T_R$.

Next, we show that the left-moving Virasoro vacuum block in the appropriate limit becomes the Schwarzian correlation function, with a  calculation closely related to \cite{Mertens:2017mtv}. At low left-moving temperature, the conformal blocks are simplest in a `direct' channel, where the OPE coefficients can be interpreted as matrix elements of intermediate states in the thermal trace, but we must compute the vacuum block in a dual channel related by the modular transformation. To relate these two channels, we use the sequence of modular S and fusion transformations pictured in figure \ref{fig:sequenceLMVB}, and the associated kernels can then be interpreted in terms of the density of states and matrix elements of light operators in the Schwarzian theory. These Schwarzian data in fact follow from a particular limit of the universal density of states and OPE coefficients recently discussed in \cite{Collier:2019weq}.

These considerations extend to higher-point functions and general time-orderings, discussed in section \ref{sec:generalcorr}, giving formulas analogous to \eqref{eqCFTSch} in appropriate ranges of kinematics. In particular, correlation functions of partial waves are always given by the Schwarzian. This applies to out-of-time order four-point functions (OTOC), implying that the Lyapunov exponent in this limit saturates the chaos bound of \cite{Maldacena:2015waa}. This is the first calculation of OTOC in 2d CFT in a fully controlled approximation, thanks to the large $T_R$ limit.

In this work, we assume the absence of any conserved current beyond Virasoro. With knowledge of the appropriate representation theoretic objects (the modular S-matrix and fusion kernel), our methods could be generalized to extended symmetry algebras (including supersymmetric CFTs). This would enhance the Schwarzian theory by an extra mode associated to the additional symmetry. For the case of an extra $U(1)$, this is analogous to the IR theory that appears in complex SYK \cite{Davison:2016ngz}. One could also apply the same methods to study near-extremal states with a large charge associated to this extra symmetry. We leave such generalizations for future work.

\subsection{Near-extremal BTZ black holes}

In AdS$_3$/CFT$_2$, our results have a dual interpretation in terms of large, near-extremal rotating BTZ black holes. In section \ref{sec:BTZ} we will perform the bulk analysis for Einstein gravity coupled to light matter fields. As stressed in the previous section, our 2d CFT results are valid even beyond holographic theories. The underlying reason is that, while the Schwarzian mode is strongly interacting, the other bulk interactions are suppressed by an additional scale, namely the very large horizon radius. Therefore, even an effective bulk dual with a cutoff on the AdS scale or larger (for example, from strings with AdS scale tension) is useful for describing physics at these very long distances. This is very much in the spirit of previous results recovering AdS long-distance physics from generic CFTs \cite{Fitzpatrick:2012yx,Fitzpatrick:2014vua,Collier:2018exn,Maxfield:2019hdt}, but rather more striking because a strongly interacting mode remains.

We discuss the details of this gravitational dual in section \ref{sec:BTZ}. We study this from the perspective of a Kaluza-Klein (KK) reduction of 3d gravity (initially pure gravity, later adding matter), which leads to an Einstein-Maxwell-dilaton theory. One may initially wonder whether the KK gauge field has a significant effect on the physics; we deal with it by working with boundary conditions describing an ensemble with fixed angular momentum (the charge dual to the KK gauge field), in which case the gauge field can (in the absence of charged matter) be integrated out to give a \emph{local} effective action, adding a term to the dilaton potential. These are rather unconventional boundary conditions from the three-dimensional point of view, since they allow a parameter of the boundary metric to fluctuate, but the usual boundary conditions are recovered simply by changing back to the ensemble with chemical potential for angular momentum. Charged matter (that is, with nonzero angular momentum, which includes KK modes) does not significantly affect this if the angular momentum carried is small compared to that of the black hole. 

The resulting theory of dilaton gravity (first written down by \cite{Achucarro:1993fd}) is in the class of Almheiri-Polchinski models \cite{Almheiri:2014cka}, and admits a nAdS$_2$ JT gravity regime, which describes the near-horizon region of rotating BTZ black holes. At low temperature, this gravitational physics becomes strongly coupled, and is described by the Schwarzian theory \cite{Maldacena:2016upp}. The near-horizon region governed by JT gravity has a parametrically large overlap with a region far from the horizon where gravity is classical, so the physics is well-described by propagation on a fixed background. Matching these two regimes allows us to recover correlation functions of the dual CFT from the asymptotic boundary of AdS$_3$.

Importantly, interactions (including those with KK modes) are suppressed in the limit of a large black hole, even for matter which is originally strongly interacting in AdS$_3$. This is the reason why the reduction to two-dimensional gravitational physics is useful: the KK modes become decoupled, independent free fields. Note that this is different from the usual case of small transverse dimensions in KK reductions, where the higher KK modes can be ignored because they become very massive.

We work in a second order metric formalism due to the presence of matter, in analogy to the higher dimensional cases studied recently in \cite{Almheiri:2016fws, Nayak:2018qej, Moitra:2019bub, Castro:2018ffi}. For pure 3d Einstein gravity, there is a more direct route to the Schwarzian, making use of the Chern-Simons formulation \cite{Witten:2007kt}, which describes perturbations around a given geometry. In this case, it is possible to completely reduce the bulk theory to a set of left- and right-moving boundary modes \cite{Cotler:2018zff}. The dynamics of these two modes is described by left- and right-moving copies of the Alekseev-Shatashvili action \cite{Alekseev:1990mp}. In the near extremal limit, the Alekseev-Shatashvili theory is known to reduce to the Schwarzian \cite{Mertens:2017mtv, Mertens:2018fds, Cotler:2018zff}. This boundary theory was previously proposed as a Goldstone mode of spontaneously broken conformal invariance for any chaotic 2d CFT \cite{Turiaci:2016cvo}. From this perspective, the Schwarzian emerges from a reparameterization mode of CFT$_2$, with dynamics governed by the conformal anomaly \cite{Haehl:2018izb}.

While our results bear some resemblance to previous considerations of near-extremal BTZ such as \cite{Balasubramanian:2003kq, Balasubramanian:2009bg}, there are crucial differences. In particular, the small but nonzero temperature is necessary; at very low temperatures, the Schwarzian becomes increasingly strongly coupled, and at exponentially low temperatures, when there is no longer a parametrically large number of available states, nonperturbative corrections from different topologies become uncontrolled \cite{Saad:2019lba}.\footnote{Aspects of such nonperturbative effects in 3D gravity will be explored in forthcoming work \cite{nonpert}.}

\section{Invitation: the near-extremal spectrum of BTZ and CFT$_2$}\label{sec:inv}

We begin by studying the partition function of near-extremal rotating BTZ black holes, and compare to the Schwarzian theory. We then obtain the corresponding spectrum directly from a very general class of CFTs, using an argument which we will later generalize to Schwarzian correlation functions. This simple example illustrates the main ideas used for the more technical calculations in section \ref{sec:CFT}.
 
\subsection{The BTZ partition function}

Our starting point is the BTZ black hole in three-dimensional pure Einstein gravity. In Euclidean signature, this is a solution whose asymptotic boundary is a torus parameterized by angle $\varphi$ and Euclidean time $t_E$, where we periodically identify $t_E$ with inverse temperature $\beta=T^{-1}$ and twist angle $\theta$:
\begin{equation}
	ds^2 = dt_E^2 + d\varphi^2,\qquad (t_E,\varphi)\sim (t_E,\varphi+2\pi)\sim (t_E+\beta,\varphi+\theta)
\end{equation}
We have chosen units such that the spatial circle has unit radius, and a corresponding dimensionless time $t_E$, so $\beta$ and energy will also be dimensionless. The Euclidean BTZ black hole is a saddle-point contribution to the dual CFT partition function on this torus; this is a grand canonical partition function, where $\theta$ plays the role of an imaginary chemical potential for the angular momentum $J$:
\begin{gather}
	\begin{aligned}
		Z(\beta,\theta) &= \Tr\left[ e^{2 \pi i \tau \left(L_0-\frac{c}{24}\right) - 2 \pi i \bar{\tau} \left(\bar{L}_0-\frac{c}{24}\right)} \right] \\
		&= \Tr\left[e^{-\beta H - i\theta J}\right]
	\end{aligned} \\
	\tau = \frac{\theta+i\beta}{2\pi}, \quad \bar{\tau} =\frac{\theta-i\beta}{2\pi}\\
	 H= L_0+\bar{L}_0-\tfrac{c}{12},\quad J=\bar{L}_0-L_0
\end{gather}
The $2\pi$ periodicity of $\theta$ is equivalent to integer quantization of $J$.

Now, the Euclidean BTZ black hole is a solid torus, where the Euclidean time circle is contractible in the bulk. The one-loop partition function of this geometry \cite{Giombi:2008vd} (which is in fact exact to all loops, up to possible renormalisation of $c$, see \cite{Maloney:2007ud} and more recently \cite{Cotler:2018zff}) can be written as
\begin{equation}
	Z_\text{BTZ} = \chi_\id(-1/\tau) \chi_\id(1/\bar{\tau})
\end{equation}
where $\chi$ is the Virasoro character of the vacuum representation,
\begin{equation}\label{eq:vacChar}
\chi_\id(\tau) =  \frac{(1-q)q^{-\frac{c-1}{24}}}{\eta(\tau)},\quad q=e^{2\pi i \tau}.
\end{equation}
To explain this result, we first note that the modular transform $\tau\mapsto -1/\tau$ swaps space and Euclidean time directions, after which we can interpret the BTZ solution as empty AdS$_3$, periodically identified with a twist. After this reinterpretation, the partition function counts perturbative excitations of AdS$_3$, which are the boundary gravitons, with dual CFT interpretation as the Virasoro descendants of the vacuum state. The central charge $c$ appears as the Casimir energy of the vacuum, which (at tree level) takes the Brown-Henneaux \cite{Brown:1986nw} value  $c=\frac{3\ell_3}{2G_N}$, where $\ell_3$ is the AdS length (the subscript distinguishing it from the two-dimensional AdS length which we will encounter later).

\subsection{The near-extremal limit}\label{PFNEL}

To recover the Schwarzian theory, we now wish to take an extremal limit, which requires low temperature (of order $c^{-1}$), with spin $J$ of order $c$ at least (and, as we will see later, much larger still for the simplest two-dimensional description). This requires a real chemical potential for the spin, corresponding to imaginary $\theta$; the corresponding Euclidean solution is then complex, while the Lorentzian solution (given explicitly in equation \eqref{eq:BTZmetric}) is real. For such a situation, we parameterize the partition function using separate left- and right-moving temperatures,
\begin{equation}
	\tau = \frac{i \beta_L}{2\pi}, \quad \bar{\tau} = -\frac{i \beta_R}{2\pi} \implies Z=  \Tr\left[ e^{-\beta_L \left(L_0-\frac{c}{24}\right) - \beta_R \left(\bar{L}_0-\frac{c}{24}\right)} \right],
\end{equation}
so that $\beta=\tfrac{1}{2}(\beta_L+\beta_R)$, $\theta = \frac{1}{2i}(\beta_R-\beta_L)$.

To approach extremality, we will take low left-moving temperature, with $\beta_L$ of order $c$ and $c\gg 1$. We will also see that it is simplest to take a very large black hole, which here means very high right-moving temperature, $\beta_R\ll 1$. In particular, for the black hole to dominate over the vacuum in the grand canonical ensemble, we require $\beta_L\beta_R<(2\pi)^2$, which constrains $\beta_R$ to be of order $c^{-1}$ or smaller.
\begin{equation}\label{eq:NEGCE}
	\text{Near extremal limit (grand canonical):}\quad c\gg 1, \quad \beta_L \text{ of order } c, \quad \beta_R \lesssim c^{-1}
\end{equation}
We can now evaluate the BTZ partition function in this limit, simply taking low temperature for the left-moving character and high temperature for right-moving:
\begin{align}
\chi_\id\left(\frac{2\pi i}{\beta_L}\right) &\sim 2\pi \left(\frac{2\pi}{\beta_L}\right)^{3/2} \exp\left[\frac{\beta_L}{24}+\frac{c}{24}\frac{(2\pi)^2}{\beta_L}\right] \\
\chi_\id\left(\frac{2\pi i }{\beta_R}\right) &\sim \exp\left[\frac{c}{24} \frac{(2\pi)^2}{\beta_R}\right]
\end{align}
For this, we use $\eta(\tau)\sim e^{\frac{i \pi}{12}\tau}$ as $\tau\to i\infty$, for the left-movers after the modular transform $\eta(-1/\tau) = \sqrt{-i\tau}\eta(\tau)$.

We already see the Schwarzian partition function appearing in the left-moving half, but to compare it is most convenient to first pass to a different ensemble, where we fix the temperature and spin:
\begin{equation}\label{eq:ZJ}
	Z_J(\beta) = \int_{-\pi}^{\pi} \frac{d\theta}{2\pi} e^{i\theta J} Z(\beta,\theta)
\end{equation}
This is a canonical ensemble in the Hilbert space of spin $J$ states, in which we can take an equivalent near-extremal limit:
\begin{equation}
	\text{Near extremal limit (canonical):}\quad  c\gg 1, \quad \beta \text{ of order } c,\qquad J\gg c,
\end{equation}
where taking the spin to be this large is not strictly necessary for now, but simplifies things more generally.

Inserting the BTZ partition function, we evaluate the integral taking us to fixed spin by saddle-point at large $J$.\footnote{\label{sl2zFootnote}The BTZ partition function is not periodic in $\theta$, so does not have a quantised spectrum in $J$. To fix this, one can sum over the restricted family of `$SL(2,\ZZ)$' black holes \cite{Maldacena:1998bw} related by the modular transforms that take $\theta\mapsto \theta+2n\pi$. Summing over this family is equivalent to extending the range of integration in \eqref{eq:ZJ} to all $\theta\in\RR$. This makes no difference in the saddle-point approximation.} In the near-extremal approximation, this is equivalent to fixing $\beta_L =2\beta$, and doing an inverse Laplace transform in the right-moving sector, in the variable $\beta_R = \beta+i\theta$. We can evaluate this at the saddle point $\beta_R = 2\pi \sqrt{\frac{c}{24J}} $, to get 
\begin{equation}\label{eq:NEPF}
	Z_J(\beta) \sim \frac{\pi}{2\sqrt{2}} \left(\frac{2\pi}{\beta}\right)^{3/2} \left(\frac{c}{6J^3}\right)^{1/4}\exp\left[2\pi\sqrt{\frac{c}{6}J}-\beta J+\frac{\beta}{12}+\frac{c}{12}\frac{\pi^2}{\beta}\right].
\end{equation}
We can match this precisely with the Schwarzian partition function:
\begin{equation}
	Z_\text{Schw}(\tilde{\beta}) = \left(\frac{\pi}{\tilde{\beta}}\right)^{3/2} e^{\pi^2/\tilde{\beta}}
\end{equation}
Our notation for Schwarzian correlation functions reflects the fact that they depend on temperature only through the combination $\tilde{\beta}=\frac{\beta}{2C}$ (and later, time $\tilde{t}=\frac{t}{2C}$). We can now write \eqref{eq:NEPF} as follows:
\begin{align}
Z_J(\beta) &= e^{S_0 - \beta E_0} Z_\text{Schw}(\tilde{\beta} = \tfrac{1}{2C}\beta)	 \\
 C &= \frac{c}{24}\\
 S_0 &\sim 2\pi \sqrt{\frac{c}{6}J}\\
 E_0 &= J-\frac{1}{12}
\end{align}
Recall that the Schwarzian coupling $C$ has dimensions of time, and the scale here is set by the radius of the circle on which we put the CFT. For large $c$, the characteristic time of the Schwarzian theory is therefore parametrically long compared to the time it takes to encircle the spatial circle.

The (temperature independent part of the) prefactor contributes a logarithmic correction to the entropy $S_0$, which for large spin can be written as $S_0 \to S_0 - \frac{3}{2} \log S_0$. Precisely this logarithmic corrections was previously studied in references \cite{Kaul:2000kf,Carlip:2000nv}. As stressed in \cite{Charles:2019tiu} this correction is important for a precise comparison between microscopic and macroscopic black hole entropy calculation.

So far we matched the partition function of a near extremal BTZ at fixed angular momentum with the Schwarzian partition function. We could try to do the same at fixed chemical potential. In this case, as we will see later, a $U(1)$ gauge mode besides the Schwarzian is relevant (the boundary conditions corresponding to fixed $J$ will eliminate the dynamics of the gauge field). Including this mode one could do the match directly in the grand canonical ensemble. In any case, the difference only arises in one-loop corrections to the partition function, and for correlation functions the ensembles are equivalent to leading order at large $J$.

Finally, it is possible to extend this analysis to the case of a gravity theory with different left and right moving central charges $c_L\neq c_R$. The argument above works out in the same way, and the partition function will be the same as the Schwarzian theory with coupling $C=c_L/24$, while the extremal entropy goes as $S_0 \sim \sqrt{c_R J}$. This can be reproduced by considering a three dimensional black hole with an additional gravitational Chern Simons term in the bulk action \cite{Verlinde:1989ua}. A similar setup, and its relation to the Schwarzian theory, was recently considered in \cite{CastroBeatrix}.

\subsection{General irrational CFTs}

Having recovered the density of states of the Schwarzian theory from near-extremal BTZ, we will now see that it is simple to recover this result for a very general class of CFTs. We give a two step argument, presented in a way that will later generalize to correlation functions. First, we show that the near-extremal partition function reduces to a vacuum character in a modular transformed expansion. Secondly, we will show that this character can be rewritten in the original channel in terms of the Schwarzian density of states.

\subsubsection{Dominance of the dual vacuum character}

For our first step, we use modular invariance of the theory, writing the partition function as a trace over a Hilbert space quantizing on a Euclidean time circle (rather than spatial circle) of the torus:
\begin{equation}
	\begin{aligned}
	Z(\beta_L,\beta_R) &= Z\left(\frac{(2\pi)^2}{\beta_L},\frac{(2\pi)^2}{\beta_R}\right) \\
	 &= \chi_\id\left(\frac{2\pi i}{\beta_L}\right) \chi_\id\left(\frac{2\pi i}{\beta_R}\right) + \sum_\text{primaries} \chi_h\left(\frac{2\pi i}{\beta_L}\right) \chi_{\bar{h}}\left(\frac{2\pi i}{\beta_R}\right)
	\end{aligned}
\end{equation}
In the second line, we have written the trace as a sum over representations of the Virasoro symmetry, introducing the characters
\begin{equation}
\chi_h(q) = \frac{q^{h-\frac{c-1}{24}}}{\eta(\tau)}
\end{equation}
of nondegenerate representations of lowest weight $h$. To write this, we have assumed that there are no currents in the theory (that is, operators with $h=0$ or $\bar{h}=0$) besides the identity representation: if there are currents, we should classify representations according to the extended algebra, and would expect to recover a Schwarzian theory with corresponding additional symmetries. For simplicity, we assume something slightly stronger, namely that there is a `twist gap' $\bar{h}_\text{gap}$, which is a positive lower bound on $\bar{h}$ for all non-vacuum primaries.

Given such a theory with large central charge and a (not necessarily large) twist gap, we can take the near-extremal limit \eqref{eq:NEGCE}, and find that the vacuum representation (in the modular transformed channel) dominates the partition function:
\begin{equation}\label{eq:nonVacSuppression}
	\frac{Z(\beta_L,\beta_R)}{Z_\text{BTZ}(\beta_L,\beta_R)} = 1+O\Big(e^{-\frac{(2\pi)^2}{\beta_R} \bar{h}_{\text{gap}}}\Big)
\end{equation}
The exponential suppression comes from the ratio of non-vacuum and vacuum right-moving characters, $\chi_{\bar{h}}\left(\frac{2\pi i}{\beta_R}\right)/\chi_\id\left(\frac{2\pi i}{\beta_R}\right)$. While this is the parametric suppression in $\beta_R$ (guaranteed for any given theory by uniform convergence of the partition function), it should be borne in mind that it can be accompanied by a prefactor which is parametrically large in $c$, so we need to take $\beta_R$ correspondingly small. We give two examples.

For the contribution of a single primary operator, the ratio of left-moving characters $\chi_{h}\left(\frac{2\pi i}{\beta_L}\right)/\chi_\id\left(\frac{2\pi i}{\beta_L}\right)$ contributes a factor of $\beta_L$, which is of order $c$, arising because the factor of $1-q$ in \eqref{eq:vacChar} subtracting null states from the vacuum. For a single state to be suppressed relative to the vacuum, we therefore must take $\beta_R \ll \frac{\bar{h}}{\log c}$. A gravitational explanation for this is that the BTZ extremality bound receives a one-loop correction of order $\exp\left(-\frac{(2\pi)^2}{\beta_R} \bar{h}\right)$ \cite{Maxfield:2019hdt}\cite{Benjamin:2019stq}, which to be ignored must be much smaller than the typical energies (of order $c^{-1}$) that we are interested in. One could perhaps account for such corrections by absorbing them into a shift of $E_0$.

For pure gravity, the twist gap $\bar{h}_\text{gap}$ is of order $c$, but there are exponentially many states close to the twist gap, so corrections in \eqref{eq:nonVacSuppression} are accompanied by an exponentially large prefactor. These states are black holes in the modular dual channel we are using for our expansion; in the direct channel, they come from thermal AdS. This is simply the Hawking-Page transition we have already encountered, requiring us to take $\beta_R \lesssim c^{-1}$ for black holes to dominate the grand canonical ensemble.

We emphasize that, while these two examples may not exhaust all possible corrections from non-vacuum characters, for any given theory there is always some sufficiently small $\beta_R$ to guarantee the dominance of the vacuum character. Our conclusions will apply to any CFT with large $c$ and a twist gap, even if its bulk dual description (if any exists) is stringy or nonlocal on the AdS scale; we may just have to take the spin of the states we study to be very large.

\subsubsection{Schwarzian spectrum from modular S matrix}\label{sec:PFCFT}

We have already shown how to recover the Schwarzian partition function from the modular transform of the vacuum character, phrased as the BTZ partition function. However, our method, which required a simple closed form expression for the vacuum character, will not be available to us for the generalization to correlation functions. We therefore find it instructive to recover the Schwarzian in a different way.

Firstly, we can treat the right-moving sector, which is very simple. We are taking a very high right-moving temperature, which corresponds to very low temperature in the modular transformed channel. This simply projects us onto the vacuum, and we are sensitive only to the Casimir energy:
\begin{equation}
	\chi_\id\left(\frac{2\pi i}{\beta_R}\right) \sim \exp\left(\frac{c}{24}\frac{(2\pi)^2}{\beta_R}\right)
\end{equation}
As we saw earlier, when we go to an ensemble of fixed spin this provides the zero temperature entropy $S_0$ and shifts the ground state energy $E_0$.

The interesting part, where the Schwarzian theory lives, is in the left-moving sector. Since the left-moving temperature is very low, the characters are simple in the `direct' expansion rather than the modular transform; they simply become the Boltzmann weights of the lowest-weight state, since the low temperature suppresses all descendants:
\begin{equation}
	\chi_h\left(\frac{i\beta_L}{2\pi}\right) \sim e^{-\beta_L\left(h-\tfrac{c}{24}\right)}
\end{equation}
We therefore directly get the density of states of the Schwarzian theory if we can decompose the modular transformed vacuum character $\chi_\id\left(\frac{2\pi i}{\beta_L}\right)$ into `direct' characters $\chi_h\left(\frac{i\beta_L}{2\pi}\right)$ (even though the context is slightly different, this argument is much the same as the one used in \cite{Mertens:2017mtv} to compute the Schwarzian partition function). 

This operation is the definition of the modular S-matrix, which we introduce presently. For this, we will use alternative parameters for central charge $c$ and operator dimension $h$:
\begin{gather}\label{eq:parcft}
	c=1+6Q^2,\quad Q=b^{-1}+b \\
	h= \frac{c-1}{24} + P^2 \quad \text{or }\quad  h=\alpha(Q-\alpha), \text{ where } \; \alpha=\tfrac{Q}{2}+iP \label{eq:parcft2}
\end{gather}
These parameters are perhaps most familiar from Coulomb gas or Liouville theory, where $Q$ is a background charge and $b$ the Liouville coupling, and $P$ is a target space momentum, but their appearance here is explained by something more universal, namely that they are the natural parameters for Virasoro representation theory. Note that there is a degeneracy of these new labels, since $h$ is invariant under the reflection $P\to -P$.

We are interested in theories at large $c$, and two ranges of operator dimensions will turn out to be important for us:
\begin{align}
	\text{Limit }& b\to 0 \implies c\to\infty \\
	\text{Schwarzian states:}&\quad k = b^{-1}P \text{ fixed }\implies h- \frac{c-1}{24}\sim \frac{6}{c} k^2 \label{eq:SchwStates} \\
	\text{Schwarzian operators:}&\quad h \text{ fixed }\implies \alpha \sim b h
\end{align}
As indicated, dimensions corresponding to fixed $k$ will turn out to correspond to energies $k^2$ in the Schwarzian limit, while fixed $h$ (which means imaginary momentum $P\sim i\frac{Q}{2}$) will correspond to operators we can insert in that limit. 

As a side comment, in the analysis of Ponsot and Teschner \cite{Ponsot:1999uf} the parameter $P$ labels unitary continuous representations of $\mathcal{U}_q(\mathfrak{sl}(2))$. As we will see below, in the Schwarzian limit the parameter $k=b^{-1} P$ will become a label of principal unitary series representations of $\mathfrak{sl}(2)$ with spin $j=\frac{1}{2} + i k$, while the dimension of these operators is related to the Casimir of these representations. We will see below how the Schwarzian limit of Virasoro is controlled by classical $\mathfrak{sl}(2)$ quantities \footnote{This limit is different than finite $L \sim  b P$ with $b\to0$ which corresponds to the classical limit of quantized Teichmuller space in the length basis. In CFT language this is called sometimes the semiclassical limit with large $c$ and $h/c$ finite. }.

Coming back to our calculation, with the notation introduced above, we can write the desired decomposition of the modular transformed vacuum block, now labeling the representations using $P$ instead of $h$:
\begin{gather}
	\chi_\id\left(-1/\tau\right) = \int_{-\infty}^\infty \frac{dP}{2} \chi_P(\tau) \modS_{P\id} \\
	\modS_{P\id} = 4\sqrt{2} \sinh(2\pi b P) \sinh(2\pi b^{-1} P) \label{eq:idS}
\end{gather}
The factor of $\tfrac{1}{2}$ in the measure here cancels the double counting of momenta $P$ and $-P$. For completeness, we note the similar decomposition of a nondegenerate block \footnote{This relation is useful to compute the exact path integral of the Schwarzian theory over different Virasoro coadjoint orbits \cite{Mertens:2019tcm}.}:
\begin{gather}
	\chi_{P'}\left(-1/\tau\right) = \int_{-\infty}^\infty \frac{dP}{2} \chi_P(\tau) \modS_{PP'} \\
	\modS_{PP'} = 2\sqrt{2} \cos(4\pi P' P)
\end{gather}
The vacuum result can be recovered from this by subtracting the null states descending from the $h=1$ state ($\chi_\id = \chi_{P=\frac{i}{2}(b^{-1}+b)}- \chi_{P=\frac{i}{2}(b^{-1}-b)}$, and similarly for the S-matrix). The characters are simple enough that these relations can be verified directly, but for the more complicated cases encountered later we will have access to the analogue of the S-matrix, but not the analogue of the characters (the conformal blocks).\footnote{This is not surprising if we take the perspective that the S-matrix and its analogues are natural representation theoretic objects. Here, the identity S-matrix $\modS_{P\id}$ is the Plancherel measure of the quantum group $\mathcal{U}_q(\mathfrak{sl}(2))$ closely associated with Virasoro \cite{Ponsot:1999uf}.}

We now finally take the near-extremal limit $c\to \infty$ with $\beta_L$ of order $c$. The characters $\chi_P(\tau)$ become Boltzmann weights as discussed earlier, and the integral is dominated by weights with $P$ of order $b$, labeled by $k$:
\begin{equation}
	\chi_\id\left(\frac{2\pi i}{\beta_L}\right) \sim e^{\tfrac{1}{24}\beta_L} 2^{3/2}(2 \pi b^3) \int_0^\infty\! d(k^2)  \sinh(2 \pi k) \,e^{-\beta_L b^2 k^2}
\end{equation}
In this integral, we recognize the Schwarzian density of states, going as $\sinh$ of the square root of energy. The prefactors contribute to $S_0$ and $E_0$. Doing the integral explicitly we can check that
\begin{gather}
	\chi_\id\left(\frac{2\pi i}{\beta_L}\right) \sim 4\pi\sqrt{2} b^3 e^{\frac{1}{24}\beta_L} Z^{(\text{Schw})}\left(\tilde{\beta}=b^2\beta_L\right),
\end{gather}
which obviously matches with the calculation done in the previous section. 

From this method, the modular S-matrix $\modS_{P\id}$ very directly gives us the spectral data of the Schwarzian theory. As anticipated, this happens to be the Plancherel measure of the universal cover of classical $\mathfrak{sl}(2)$ if $k$ is interpreted as the label of the principal series. To understand this, we recall the close relation between the Schwarzian and $\mathfrak{sl}(2)$ through a BF formulation of JT gravity \cite{Blommaert:2018oro}\cite{Iliesiu:2019xuh} (see also \cite{Kitaev:2018wpr} for another interpretation of the relation with $\mathfrak{sl}(2)$). Besides the principal series, the discrete series representation of $\mathfrak{sl}(2)$ will make an appearance when computing correlators. 

We could perform a similar analysis for the right-moving sector, noting in particular that for large $\bar{P}$ we recover the `Cardy' density of states $S_0= 2\pi \sqrt{\frac{c}{6}J}$ from that limit of $\modS_{\id \bar{P}}$. However, this is not simple or natural, because the descendant states are very important at the high right-moving temperatures of interest. In this example, we cannot read $S_0$ directly from the modular S-kernel, because it counts only primaries, and is insensitive to the contamination from descendants, leading to a discrepancy in the logarithmic corrections. Our logic will always be to `project onto the vacuum' in the right-moving sector, and then perform the modular transform to find the spectral data of the left-moving sector, where the Schwarzian resides.

\section{CFT correlation functions}\label{sec:CFT}

In this section, we discuss the correlation functions of CFTs in the near-extremal limit of the grand canonical ensemble $\beta_R\to 0$ with $\beta_L$ of order $c$, under the previously introduced conditions of a twist gap and large $c$. Note that we do not require the CFT to be `holographic', because additional assumptions of a sparse light spectrum or 't Hooft factorization are unnecessary (though our results could be strengthened under such assumptions). We obtain Lorentzian correlation functions at time separations $t$ of order $c$, including dependence on angular separations $\varphi$. For comparison to the two-dimensional dual gravitational physics in the next section, we extract the S-wave correlation functions, where the operators are averaged over the circle.

We focus mainly on the simplest correlation function of interest, namely the two-point function of identical operators in the near-extremal limit \eqref{eq:NEGCE} of the grand canonical ensemble. This will be sufficient to illustrate the main ideas, and we will discuss the general case in section \ref{sec:generalcorr}. Our method parallels that of section \ref{sec:PFCFT}: we first choose a conformal block decomposition of the correlation function such that the large spin limit $\beta_R\to 0$ selects only the vacuum block. The right-moving block is then simple to evaluate, but the left-moving block is more complicated. We follow the methods of \cite{Mertens:2017mtv} to show that this block contains the Schwarzian correlation function in the appropriate limit, by reexpressing it in a new channel which makes manifest the matrix elements of the operators with intermediate states.

For a given primary operator $\op$ with conformal weights $(h,\bar{h})$, we consider its two-point Wightman function (we will discuss time ordering and extract the retarded correlators later):
\begin{equation}
	\langle \op(z,\bar{z})\op(0,0)\rangle_{\beta_L,\beta_R} =  \Tr\left[ \op(z,\bar{z})\op(0,0) e^{-\beta_L \left(L_0-\frac{c}{24}\right) - \beta_R \left(\bar{L}_0-\frac{c}{24}\right)} \right]
\end{equation}
For Lorentzian kinematics, the coordinates $z,\bar{z}$ are lightcone coordinates on the Lorentzian cylinder
\begin{equation}\label{EqZZbar}
	z=\varphi -t,\quad \bar{z} = \varphi + t,
\end{equation}
so in particular are $2\pi$ periodic $(z,\bar{z})\sim (z+2\pi,\bar{z}+2\pi)$. In imaginary time $t_E=it$ they become complex coordinates on the torus, so additionally have the `KMS' periodicity $(z,\bar{z})\sim (z+2\pi\tau,\bar{z}+2\pi\bar{\tau})$, where $\tau=i\tfrac{\beta_L}{2\pi}$ and $\bar{\tau}=-i\tfrac{\beta_R}{2\pi}$ \footnote{Since we are taking $\beta_R$ very small and $\beta_L$ very large this identification makes $z$ approximately periodic and is reminiscent of the DLCQ limit \cite{Balasubramanian:2009bg}. We will not use this perspective in this paper though. }. The operator ordering is determined by ordering in Euclidean time, provided in Lorentzian kinematics by an appropriate $i\epsilon$ prescription giving an imaginary part to $t$.

\subsection{Right-moving sector: dominance of a vacuum block}\label{sec:rmsdvb}

The high right-moving temperature allows us to simplify the correlation function in the right-moving sector, by decomposing the amplitude with respect to quantization by spatial translations, instead of time evolution. Translations then suppress any intermediate state with $\bar{h}>0$. We can phrase this as doing a modular transform to make the right-moving temperature very low, before writing the thermal trace as a sum over states, as well as inserting a complete set of intermediate states between the two operator insertions.

This means we are decomposing the correlation function into conformal blocks by inserting a complete set of states along a pair of Euclidean time cycles, separating the two insertions of $\op$. This is a `necklace' channel decomposition of the correlation function, which we label by $\widetilde{N}$, with the tilde indicating that the intermediate states are inserted on Euclidean time circles, not spatial circles.
Explicitly, we can write the conformal block expansion in this `$\widetilde{\text{Necklace}}$' channel as follows:
\begin{gather}\label{eq:necklaceBlock1}
	\langle \op(z,\bar{z})\op(0,0)\rangle_{\beta_L,\beta_R} = \sum_{\substack{\text{primaries} \\ \op_1,\op_2}} |C_{\op\op_1\op_2}|^2\mathcal{F}^{(\tilde{N})}_{h_1,h_2}(z,\beta_L) \bar{\mathcal{F}}^{(\tilde{N})}_{\bar{h}_1,\bar{h}_2}(\bar{z},\beta_R) \\
	\begin{aligned}
	\bar{\mathcal{F}}^{(\tilde{N})}_{\bar{h}
	_1,\bar{h}_2}(\bar{z},\beta_R) &= \left(\frac{2\pi}{\beta_R}\right)^{2\bar{h}}\sum_{N_1,N_2} \langle \bar{h}_2,N_2 |\op |\bar{h}_1,N_1\rangle \langle \bar{h}_1,N_1|\op|\bar{h}_2,N_2\rangle \\
	&\qquad \times\exp\left[-\bar{z}\tfrac{2\pi}{\beta_R}(\bar{h}_1+|N_1|-\tfrac{c}{24})-(2\pi-\bar{z})\tfrac{2\pi}{\beta_R}(\bar{h}_2+|N_2|-\tfrac{c}{24})\right]
	\end{aligned}\nonumber
\end{gather}
We have written the block decomposition in the `barred' right-moving sector, which is most relevant for our present considerations; there is a similar expression for the left-moving blocks. The states $|\bar{h},N\rangle$ are an orthonormal basis of descendants (at level $|N|$) of a primary state with weight $\bar{h}$. In the expression for the conformal block, we implicitly change the normalization of $\op$ to set $\langle \bar{h}_2|\op|\bar{h}_1\rangle$ to unity; this absorbs OPE coefficients, and also gives the prefactor in the block coming from the rescaling of the Euclidean time circle from length $2\pi$ to $\beta_R$. The exponential factors implement translation (generated by $\frac{2\pi}{\beta_R}(\bar{L}_0-\tfrac{c}{24})$, where the factor comes from the same rescaling of lengths), first by $\bar{z}$ between the operator insertions, and then by $2\pi-\bar{z}$ to complete the spatial circle.

Now, when we take the $\beta_R\to\infty$ limit, choosing $0<\bar{z}<2\pi$ as we always may by the periodicity $(z,\bar{z})\sim (z+2\pi,\bar{z}+2\pi)$, it is manifest in this decomposition that intermediate primaries $\op_1,\op_2$ with large $\bar{h}_1,\bar{h}_2$ and right-moving descendants are suppressed. Explicitly, the descendants drop out of the right-moving blocks, so we have\footnote{We can guarantee that the infinite sum of terms is suppressed since the sum converges uniformly for any range of $\beta_R$ bounded away from zero.}
\begin{equation}
	\bar{\mathcal{F}}^{(\tilde{N})}_{\bar{h}_1,\bar{h}_2}(\bar{z},\beta_R) \sim \left(\frac{2\pi}{\beta_R}\right)^{2\bar{h}} \exp\left[-\frac{2\pi}{\beta_R}\left(\bar{z} \bar{h}_1+(2\pi-\bar{z})\bar{h}_2-2\pi\tfrac{c}{24}\right)\right] \,.
\end{equation}
From this, the dominant contribution comes from intermediate operators that minimize $\bar{z} \bar{h}_1+(2\pi-\bar{z})\bar{h}_2$, under the condition that the OPE coefficient $C_{\op\op_1\op_2}$ is nonzero. In particular, this condition means that we cannot choose both $\op_1$ and $\op_2$ to be the identity. For sufficiently small $\bar{z}$, the dominant choice is $\op_1=\op$, $\op_2=\id$; this remains true in a finite ($\beta_R$-independent) range $0<\bar{z}<\bar{z}_*$, where $\bar{z}_*$ is lower bounded by $2\pi\frac{\bar{h}_\text{gap}}{\bar{h}}$ (or $\bar{z}_*=\pi$ if that is smaller, from $\op_1=\id$, $\op_2=\op$). For now, we will fix $\bar{z}$ in this range; this is the most important region (in particular, dominating partial waves for any fixed angular momentum $\ell$), since the correlator is exponentially suppressed when $|\bar{z}|$ is not small.

This behavior has a simple gravitational interpretation, which is easiest to state for spacelike Wightman functions. The two-point function gives an amplitude for a spacelike propagation of some particle between the insertion points, going a long distance because there is a very large black hole in the way. The particle can choose to go either way around the black hole, corresponding to the choices $\op_1=\op$ and $\op_2=\id$, or $\op_1=\id$ and $\op_2=\op$. However, the amplitude may be larger for the particle to split into two lighter particles (dual to $\op_1,\op_2$), each going a different way around the black hole, before rejoining on the far side.\footnote{The corresponding amplitude with both particles on the same side is not relevant, since it has already been absorbed as a `vacuum polarization' renormalization of $\op$, by choosing $\op$ to be a primary of definite scaling dimension: since BTZ is locally isometric to AdS$_3$, the only physical corrections come from configurations that make use of the nontrivial topology of the spacetime.}

For $0<\bar{z}<\bar{z}_*$, we therefore have dominance of this particular vacuum block, just as the vacuum character dominated the partition function in equation \eqref{eq:nonVacSuppression}. The right-moving block becomes very simple in the $\beta_R\to 0$ limit, reducing to the exponential we have seen already if we fix $\bar{z}$ of order one. In fact, we can do a little better, giving an answer that works also for small $\bar{z}$ of order $\beta_R$ or less, when the descendants of $\op$ in the intermediate states are not negligible. The descendants are suppressed in the sum over $N_2$, so the only relevant state is the vacuum, which is equivalent to replacing the torus with an infinite cylinder. The block therefore becomes the well-known result for the two-point function at finite temperature on an infinite line.

We can summarize the results of this section as follows:
\begin{gather}\label{eq:Ntildedominance}
	\langle \op(z,\bar{z})\op(0,0)\rangle_{\beta_L,\beta_R} \sim \mathcal{F}^{(\tilde{N})}_{\op,\id}(z,\beta_L) \bar{\mathcal{F}}^{(\tilde{N})}_{\op,\id}(\bar{z},\beta_R) \\
	\bar{\mathcal{F}}^{(\tilde{N})}_{\op,\id}(\bar{z},\beta_R)\sim e^{\frac{(2\pi)^2}{\beta_R}\frac{c}{24}} \left[\frac{\beta_R}{\pi}\sinh\left(\frac{\pi}{\beta_R}\bar{z}\right)\right]^{-2\bar{h}}
\end{gather}
So far, this applies for any theory with a twist gap (with any central charge) in the $\beta_R\to 0$ limit, either scaling $\bar{z}\propto \beta_R$ or with fixed $0<\bar{z}<\bar{z}_*$. Our remaining task is to determine the left-moving identity block $\mathcal{F}^{(\tilde{N})}_{\op,\id}(z,\beta_L)$ in the limit of interest, taking $c\to\infty$ while keeping $z$ and $\beta_L$ proportional to $c$.

\subsection{Left-moving sector: the Schwarzian}\label{sec:leftmovblock2pt}

Much like in the previous section, there is a preferred channel in which the left-moving blocks simplify, now because we are taking low left-moving temperature $\beta_L\propto c\gg 1$. This corresponds to performing the usual quantization by time evolution, inserting complete sets of states in the thermal trace and between operator insertions. Just as in \eqref{eq:necklaceBlock1}, we can write the blocks in this `direct' necklace channel, which we label by $N$, as follows\footnote{The phase arises because the $z$ coordinate is rotated by $\frac{\pi}{2}$ relative to the time evolution, giving a factor $e^{i\frac{\pi}{2}h}$ for each operator insertion.}:
\begin{equation}\label{eq:necklaceBlock2}
	\begin{aligned}
	\mathcal{F}^{(N)}_{h_1,h_2}(z,\beta_L) &= e^{-i\pi h}\sum_{N_1,N_2} \langle h_2,N_2 |\op |h_1,N_1\rangle \langle h_1,N_1|\op|h_2,N_2\rangle \\
	&\qquad \times\exp\left[i z(h_1+|N_1|-\tfrac{c}{24})-(\beta_L+ i z)(h_2+|N_2|-\tfrac{c}{24})\right]
	\end{aligned}\nonumber
\end{equation}
Now, for $\Im z>0$, which corresponds to the time-ordering where the insertion of $\op(z,\bar{z})$ comes after the insertion of $\op(0,0)$, the descendants are suppressed in the limit of interest. This is a limit $c\to\infty$ with $\beta_L \propto c$, but also (as we will see later) where we take
\begin{equation}
	h_{1,2} = \frac{c-1}{24}+\frac{6}{c}k_{1,2}^2, \quad c\to \infty,\quad k_{1,2} \text{ fixed}
\end{equation}
as in \eqref{eq:SchwStates}. The OPE coefficients of descendants are then suppressed by factors of $\frac{(h_1-h_2)^2}{c}$ and $\frac{h^2}{c}$ \cite{Fitzpatrick:2014vua,Collier:2018exn}, and we have
\begin{align}
	\mathcal{F}^{(N)}_{h_1,h_2}(z,\beta_L) &\sim e^{-i\pi h} e^{i z (h_1-h_2)-\left(h_2-\tfrac{c}{24}\right)\beta_L} \\
	&\sim e^{-i\pi h} e^{\frac{1}{24}\beta_L} e^{i \frac{6z}{c} (k_1^2-k_2^2)-\frac{6\beta_L}{c}k_2^2}. \label{eq:Nblocklimit}
\end{align}

However, unlike for the right-movers, no single operator dominates in this necklace channel $N$ where the blocks are simple. Instead, the correlation function is given by a vacuum block in the $\widetilde{N}$ channel, in which the blocks are more complicated. Our strategy is to evaluate the $\widetilde{N}$ identity block by decomposing it in terms of $N$ channel blocks, which we can write simply. This conversion between different channels is implemented by fusion and modular S transformations (and, relevant later for higher-point out of time order correlators, braiding), generalizing the modular S transform we used in section \ref{sec:PFCFT}. Fortunately, these transformations are known in relatively simple, explicit closed forms.

\subsubsection*{Warm-up: four-point identity block}
As a warm-up, we first tackle a slightly simpler problem, finding a `Schwarzian limit' of the Virasoro four-point identity block. This is a microcanonical version of the calculation we are interested in, where instead of taking a low temperature limit to take us near extremality, we fix a primary state $|\psi\rangle$ with dimension in the Schwarzian range $h_\psi = \frac{c-1}{24}+\frac{6}{c}k_\psi^2$, where $k_\psi$ is fixed in the $c\to\infty$ limit. The cylinder kinematics we are using are then related to the usual four-point cross-ratio $x$ by $x=e^{i z}$, since in radial quantization, we insert the operator $\op_\psi$  creating the state $|\psi\rangle$ at the origin and infinity, and $\op$ at $1$ and $x=e^{i z}$. We then wish to compute the identity block in the `T-channel', where we take the OPE of the two insertions of $\op$.  Note that, unlike for the torus two-point function, we do not have a general argument that this identity block always dominates the four-point correlation function, which would require conditions on the OPE coefficients $C_{\psi \op t}$. Such a result would follow from the canonical result under the assumption of a version of the eigenstate thermalization hypothesis \cite{ETH}, applied to near-extremal large spin states.

To compute the T-channel identity, we will reexpress it in terms of the `S-channel' blocks, where we take the OPE between $\op_\psi(0)$ and $\op(x)$, which are simple:
\begin{align}
	\mathcal{F}_{t}\sbmatrix{\op & \op \\ \psi & \psi}(1-x) &= \sum_N  \langle \op|\mathcal{L}_{-N}\op_{h_t}|\op\rangle\langle\psi|\mathcal{L}_{-N}\op_{h_t}|\psi\rangle   \;(1-x)^{-2h+h_t+|N|} \\
	\mathcal{F}_{s}\sbmatrix{\op & \psi \\ \op & \psi}(x) &= \sum_N  \langle\psi|\op|h_s,N\rangle\langle h_s,N| \op |\psi\rangle   \;x^{-h_\psi-h+h_s+|N|} \nonumber\\
	&\sim x^{-h + \frac{6}{c}(k_s^2-k_\psi^2)}
\end{align}
In the S-channel block, we have taken the appropriate `Schwarzian' limit of operator dimensions and kinematics, in which case the descendants drop out for the same reason as before.

Now, we can evaluate the T-channel blocks if we can decompose them into S-channel blocks, since these become simple power laws in the limit of interest. Fortunately, there is an object that does precisely this, namely the fusion kernel $\fusion$ \cite{Moore:1988qv}, which has the defining property
\begin{equation}\label{eq:4ptFusion}
	\mathcal{F}_{t}\sbmatrix{\op & \op \\ \psi & \psi}(1-x) = \int_{-\infty}^\infty \frac{dP_s}{2} \mathcal{F}_{s}\sbmatrix{\op & \psi \\ \op & \psi}(x) \; \fusion_{ P_s P_t} \sbmatrix{P & P_\psi \\ P & P_\psi},
\end{equation}
where we have used the variable $P$ introduced in \eqref{eq:parcft} to label operator dimensions $h=\frac{c-1}{24}+P^2$. The fusion kernel $\fusion$ is a kinematic object associated to Virasoro symmetry and it was computed explicitly in \cite{Ponsot:1999uf}. We will not quote the most general formula since we will only need certain special cases, but it can be found for example in equations (2.10)-(2.12) of \cite{Collier:2018exn}. 

\begin{figure} [t!]
\begin{center}
\begin{tikzpicture}
\draw[thick] (0,-1.5) -- (0,1.5);
\draw[thick] (0,0) -- (1,0);
\draw[thick,blue] (1,0) -- (1.5,0.5);
\draw[thick,blue] (1,0) -- (1.5,-0.5);
\draw (1.75,0.6) node  {\small$\mathcal{O}$};
\draw (1.75,-0.6) node  {\small$\mathcal{O}$};
\draw (0.3,1.5) node  {\small$\psi$};
\draw (0.3,-1.5) node  {\small$\psi$};
\draw (0.55,0.3) node  {\small$\mathcal{O}_t$};
\draw (4,0) node  { $=\int \frac{dP_s}{2} ~ \fusion_{ P_s P_t} \sbmatrix{P & P_\psi \\ P & P_\psi}$};
\draw[thick] (6+0,-1.5) -- (6+0,1.5);
\draw[thick,blue] (6+0,0.75) -- (6+1,0.75);
\draw[thick,blue] (6+0,-0.75) -- (6+1,-0.75);
\draw (1.5+4.9,0) node  {\small $\mathcal{O}_s$};
\draw (6+0.3,1.5) node  {\small$\psi$};
\draw (6+0.3,-1.5) node  {\small$\psi$};
\draw (1.5+5.8,0.7) node  {\small$\mathcal{O}$};
\draw (1.5+5.8,-0.7) node  {\small$\mathcal{O}$};
\end{tikzpicture}
\end{center}
\vspace{-0.5cm}
\caption{\small Diagram of the fusion transformation that was used to compute the left moving vacuum block in the appropriate limit. The blue lines correspond to the two operator insertions.}
\label{fig:fusion}
\end{figure}
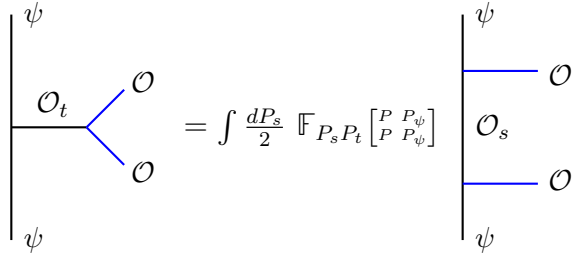

Here, we need a special case of the fusion kernel, where the T-channel representation is the identity, denoted by $\id$. In this case, the fusion kernel simplifies \cite{Vartanov:2013ima, Collier:2018exn}, and can be written as \cite{Collier:2019weq}
\begin{equation}\label{CMMid}
	\fusion_{ P_s \id} \sbmatrix{P & P_\psi \\ P & P_\psi}  = \rho_0(P_s) C_0(P_s,P,P_\psi),
\end{equation}
where $\rho_0$ and $C_0$ are universal functions, appearing as densities of states and averaged OPE coefficients in a very general class of theories and a variety of limits \cite{Collier:2019weq}. We saw $\rho_0$ already in equation \eqref{eq:idS} as the decomposition of the identity character into modular transformed characters; $C_0$ can be written in closed form in terms of a special `deformed Gamma-function' $\Gamma_b$:
\begin{align}
	\rho_0(P) &= \modS_{P\id}[\id] = 4\sqrt{2}\sinh(2\pi bP)\sinh(2\pi b^{-1}P) \label{eq:rho0} \\
	C_0(P_1,P_2,P_3) &= \frac{1}{\sqrt{2}}{\Gamma_b(2Q)\over \Gamma_b(Q)^3}\frac{\prod_{\pm\pm\pm}\Gamma_b\left(\tfrac{Q}{2}\pm iP_1\pm iP_2 \pm iP_3\right)}{\prod_{k=1}^3\Gamma_b(Q+2iP_k)\Gamma_b(Q-2iP_k)} \label{eq:C0}
\end{align}
In the numerator, we take the product over all eight possible combinations of sign choices. These objects simplify further when we take a `Schwarzian limit' of large $c$, with either $P=b k$ corresponding to near-extremal operators, or $h$ fixed. The two relevant limits are as follows:
\begin{align}
	\rho_0(b k) &\sim 8\sqrt{2} \pi b^2 k \sinh(2\pi k) \label{eq:rho0limit} \\
	C_0(b k_1,b k_2,i(\tfrac{Q}{2}-bh)) &\sim  \frac{b^{4h}}{\sqrt{2}(2\pi b)^3} \frac{\prod_{\pm\pm}\Gamma(h\pm ik_1\pm ik_2)}{\Gamma(2h)} \label{eq:C0limit}
\end{align}
This is straightforward to derive using the identities 
\beq
	\frac{\Gamma_b(n Q + b y)}{\Gamma_b(nQ)} \sim \left(\frac{\sqrt{2\pi} b^{n-1/2}}{\Gamma(n)}\right)^{y} \qquad (n>0),~~~\frac{\Gamma_b(b y)}{\Gamma_b(Q)} \sim \frac{(2\pi b^3)^{y/2}}{2\pi b} \Gamma(y)
\eeq
valid in the $b\to0$ limit with fixed $y$ and integer $n$. 

Those who have studied the Schwarzian theory will immediately find these formulas somewhat familiar \cite{Mertens:2017mtv, Kitaev:2018wpr, Yang:2018gdb}.\footnote{In Appendix \ref{app:connliouville} we make the connection with Liouville theory and the calculation of \cite{Mertens:2017mtv}.} First, as we have already seen, $\rho_0$ is proportional to the density of states of the Schwarzian theory, so we can write integrals over $P$ (if they are dominated by integrating over $P$ of order $b$) as
\begin{gather}
	\int_{-\infty}^\infty \frac{dP}{2} \rho_0(P) f(P) \sim 4\pi\sqrt{2} b^3\int_0^\infty d\mu(k) f(bk) \\
	d\mu(k) = 2kdk \sinh (2\pi k),
\end{gather}
where we have introduced the measure $d\mu(k)$ encoding the Schwarzian density of states:
\begin{equation}\label{eq:ZSchw}
	Z^{(\text{Schw})}\big(\tilde{\beta}\big) = \int d\mu(k) e^{-\tilde{\beta}k^2} = \big(\tfrac{\pi}{\tilde{\beta}}\big)^{3/2} e^{\pi^2/\tilde{\beta}}
\end{equation}
Secondly, the result for $C_0$ is proportional to matrix elements appearing  in Schwarzian correlation functions.

We can now apply these results to the four-point identity block, using \eqref{eq:4ptFusion}. First, we can justify focusing on $P_s$ of order $b$ in the intermediate channel because the fusion kernel is exponentially suppressed for larger intermediate values. Putting everything together, we find our result for the T-channel vacuum block:
\begin{align}\label{eq:SchwMicro}
	\mathcal{F}_{t}\sbmatrix{\op & \op \\ \psi & \psi}(1-x) &= \int_{-\infty}^\infty \frac{dP_s}{2} \rho_0(P_s)  C_0(P_s,P,P_\psi) \, \mathcal{F}_{s}\sbmatrix{\op & \psi \\ \op & \psi}(x) \\
	&\sim \frac{2b^{4h} x^{-h}}{(2\pi)^2}\int_0^\infty d\mu(k_s)  \frac{\prod_{\pm\pm}\Gamma(h\pm ik_s\pm ik_\psi)}{\Gamma(2h)} x^{b^2 (k_s^2-k_\psi^2)}\nonumber
\end{align}
As a check, we can evaluate this in the limit $x\to 1$, in which case the integral is dominated by large $k_s$; we find that $\mathcal{F}_{t}\sim (1-x)^{-2h}$, giving the expected short distance behavior of the block, with the usual normalization.

Before returning to the torus block, we discuss a limit of our result for the identity block, taking $k_\psi \gg 1$. In that case, the integral is dominated by $k_s$ close to $k_\psi$, with $k_s-k_\psi=\delta$ of order one:
\begin{equation}\label{eq:sc4pt}
\begin{aligned}
	\mathcal{F}_{t}\sbmatrix{\op & \op \\ \psi & \psi}(1-x) &\sim \frac{x^{-h}}{2\pi} (2b^2 k_\psi)^{2h}\int d\delta  e^{\pi\delta}\frac{\Gamma(h+ i\delta)\Gamma(h- i\delta)}{\Gamma(2h)} x^{2b^2 k_\psi \delta} \\
	&= x^{-h} \left(\frac{x^{-i b^2 k_\psi}-x^{i b^2 k_\psi}}{2i b^2 k_\psi}\right)^{-2h}
\end{aligned}
\end{equation}
This is the same result found for the vacuum block in the `heavy-light' limit \cite{Fitzpatrick:2014vua,Fitzpatrick:2015zha}, where $c$ was taken to infinity with $h_\psi/c$ fixed (but different from $\frac{1}{24}$). In fact, in \cite{Collier:2018exn} this result was derived with precisely the method used here. Our result thus interpolates smoothly to this different regime.

As anticipated, the T-channel vacuum block found in equation \eqref{eq:SchwMicro} is equal to the microcanonical two-point function of the Schwarzian theory between energy eigenstates labeled by $k_\psi$ \cite{Lam:2018pvp}. Equation \eqref{eq:sc4pt} can be understood as a statement of equivalence of canonical and microcanonical ensembles in the thermodynamic limit of the Schwarzian theory, since 
\beq\label{eq:scthermal}
\mathcal{F}_{t}\sbmatrix{\op & \op \\ \psi & \psi}(1-x) \propto \left(\frac{x^{-i b^2 k_\psi}-x^{i b^2 k_\psi}}{2 b^2 k_\psi}\right)^{-2h} = \left[ \frac{\beta_\psi}{\pi} \sinh \left( \frac{\pi}{\beta_\psi} z\right) \right]^{-2h}
\eeq
is the thermal correlator with temperature $\beta_\psi = \frac{\pi}{b^2 k_\psi}$ (the proportionality factor arising from the conformal map between the plane and the cylinder).

Note that \eqref{eq:scthermal} has a thermal KMS periodicity, which in particular leads to a singularity at $z=i\beta_\psi$ as a thermal image of the short-distance singularity. This was dubbed a `forbidden singularity' in \cite{Fitzpatrick:2016ive}, since it cannot appear in the exact four-point block or correlation function. As noted in \cite{Collier:2018exn}, it is associated with a divergence of the integral in \eqref{eq:sc4pt} as $\delta\to -\infty$, but this is an artifact of the approximation we are making in the integrand, which is valid only for $e^{-\beta_\psi}<|x|<1$ (with $k_\psi \log x$ fixed in the limit $k_\psi\to\infty$). The exact Schwarzian integral \eqref{eq:SchwMicro} does not have such a divergence, so provides a regularization which resolves the forbidden singularity.

\subsubsection*{Returning to the torus}

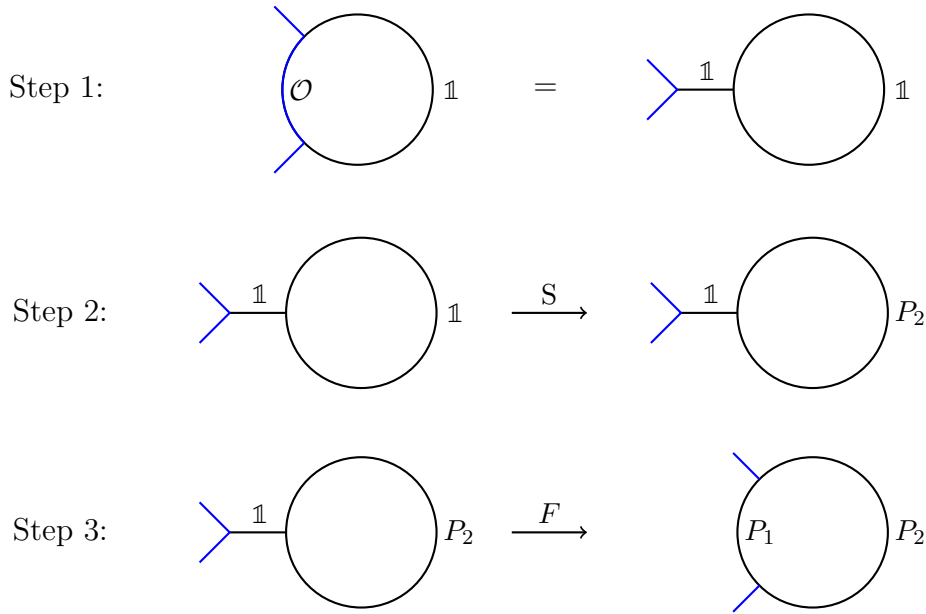
\begin{figure} [t!]
\begin{center}
\hspace{-0.25cm}\begin{tikzpicture}
\draw (-4,0) node  { Step 1:};
\draw[thick] (0,0) circle (1cm);
\draw[thick,blue] (-0.707,-0.707) arc (45:-45:-1cm);
\draw[thick,blue] (-0.707,-0.707) -- (-0.707-0.4,-0.707-0.4);
\draw[thick,blue] (-0.707,0.707) -- (-0.707-0.4,0.707+0.4);
\draw (1.25,0) node  {\small$\id$};
\draw (-0.75,0) node  {\small$\mathcal{O}$};
\draw (2.5,0) node  { $=$};
\draw[thick] (6,0) circle (1cm);
\draw[thick] (6-1,0) -- (6-1.75,0);
\draw[thick,blue] (6-1.75-0.4,0+0.4) -- (6-1.75,0) -- (6-1.75-0.4,0-0.4);
\draw (6+1.25,0) node  {\small$\id$};
\draw (6-1.35,0.25) node  {\small$\id$};
\end{tikzpicture}\\
\vspace{0.8cm}
\begin{tikzpicture}
\draw (-4,0) node  { Step 2:};
\draw[thick] (0,0) circle (1cm);
\draw[thick] (-1,0) -- (-1.75,0);
\draw[thick,blue] (-1.75-0.4,0+0.4) -- (-1.75,0) -- (-1.75-0.4,0-0.4);
\draw (+1.25,0) node  {\small$\id$};
\draw (-1.35,0.25) node  {\small$\id$};
\draw[thick,->] (2,0) -- (3,0);
\draw (2.5,0.25) node  {\small S};
\draw[thick] (6,0) circle (1cm);
\draw[thick] (6-1,0) -- (6-1.75,0);
\draw[thick,blue] (6-1.75-0.4,0+0.4) -- (6-1.75,0) -- (6-1.75-0.4,0-0.4);
\draw (6+1.3,0) node  {\small$P_2$};
\draw (6-1.35,0.25) node  {\small$\id$};
\end{tikzpicture}\\
\vspace{0.8cm}
\begin{tikzpicture}
\draw (-4,0) node  { Step 3:};
\draw[thick] (0,0) circle (1cm);
\draw[thick] (-1,0) -- (-1.75,0);
\draw[thick,blue] (-1.75-0.4,0+0.4) -- (-1.75,0) -- (-1.75-0.4,0-0.4);
\draw[thick,->] (2,0) -- (3,0);
\draw (2.5,0.25) node  {\small $F$};
\draw[thick] (6,0) circle (1cm);
\draw[thick,blue] (6-0.707,-0.707) -- (6-0.707-0.35,-0.707-0.35);
\draw[thick,blue] (6-0.707,0.707) -- (6-0.707-0.35,0.707+0.35);
\draw (+1.3,0) node  {\small$P_2$};
\draw (-1.35,0.25) node  {\small$\id$};
\draw (6+1.3,0) node  {\small$P_2$};
\draw (6-0.7,0) node  {\small$P_1$};
\end{tikzpicture}
\end{center}
\vspace{-0.5cm}
\caption{\small Diagram of the fusion and modular transformation that was used to compute the left moving vacuum torus block in the appropriate limit.  In the top diagrams, the circle is a spatial circle, and in the bottom diagrams it is a Euclidean time circle; these are swapped by the S-transform in step 2.}
\label{fig:sequenceLMVB}
\end{figure}

We now return to our discussion of the torus two-point function, for which we would like to compute the identity block in the $\widetilde{N}$ channel in \eqref{eq:Ntildedominance}. The blocks are simple in the $N$ channel \eqref{eq:Nblocklimit}, so we would like to decompose the $\widetilde{N}$ identity block in terms of $N$ blocks, with an expression analogous to \eqref{eq:4ptFusion} which expanded a T-channel block in terms of S-channel blocks for the four-point function. The idea is much the same, but we must go through some intermediate steps, following the Moore-Seiberg construction \cite{Moore:1988qv}, recently reviewed in the current context of irrational theories in \cite{Collier:2019weq}, to which we refer the reader for details. The sequence of moves we use is illustrated in figure \ref{fig:sequenceLMVB}.

First, we note that the $\widetilde{N}$ identity block is in fact equal to an identity block in a different channel, denoted $\widetilde{OPE}$. For the block decomposition in this channel, we take the OPE between the two insertions of $\op$, and insert a single complete set of states propagating in the spatial circle. In general, this is related to the $\widetilde{N}$ decomposition \eqref{eq:necklaceBlock1} by a fusion move, with external operators $\op,\op,\op_2,\op_2$ and internal operator $\op_1$ (using the labeling following \eqref{eq:necklaceBlock1}). However, for the vacuum block, we have $\op_2=\id$, so this move is trivial: in the fusion of $\id$ with itself, only $\id$ appears, so the $\widetilde{N}$ vacuum block equals the $\widetilde{OPE}$ vacuum block. For the next step, we want to replace the complete set of intermediate states propagating in the spatial circle (the $\widetilde{OPE}$ channel) with states propagating in the Euclidean time circle (the $OPE$ channel). This is a modular S-transform, with kernel $\modS$. In general, this would be the S-transform appropriate for a torus one-point function, where the external operator is determined by the representation appearing in the OPE. For us, this representation is the identity, so the modular S kernel is the same one we used for the partition function, given in  \eqref{eq:idS} and \eqref{eq:rho0}. Our final move takes us from the $OPE$ channel to the necklace $N$ channel, and is the same fusion move we just used for the four-point function; the only difference is that the external operator $\op_\psi$ is now whichever intermediate operator appears in the thermal sum over states, labeled here by $P_2$.

In equations, the moves we have just described are as follows:
\begin{align}\label{tvblm}
	\mathcal{F}^{(\tilde{N})}_{h,\id} &= \mathcal{F}^{(\widetilde{OPE})}_{\id,\id} \\
	&= \int \frac{dP_2}{2} \modS_{\id P_2} \mathcal{F}^{(OPE)}_{\id,P_2} \\
	&= \int \frac{dP_1}{2} \frac{dP_2}{2} \fusion_{\id P_1} \sbmatrix{P & P_2 \\ P & P_2} \modS_{\id P_2} \mathcal{F}^{(N)}_{P_1,P_2} \\
	&= \int \frac{dP_1}{2} \frac{dP_2}{2} \rho_0(P_1)\rho_0(P_2)C_0(P_1,P_2,P) \mathcal{F}^{(N)}_{P_1,P_2}
\end{align}
We have written the final expression in terms of our universal functions \eqref{eq:rho0}, \eqref{eq:C0}.

This expression holds in complete generality, but we can now take a limit to extract an explicit expression for $\mathcal{F}^{(\tilde{N})}_{h,\id}$. For this, we simply substitute \eqref{eq:Nblocklimit} for $\mathcal{F}^{(N)}_{P_1,P_2}$, along with \eqref{eq:rho0limit} and \eqref{eq:C0limit} for the Schwarzian limits of $\rho_0$ and $C_0$ to obtain our final result for the left-moving part of the correlation function:
\begin{gather}\label{eq:leftmovingres}
	\mathcal{F}^{(\tilde{N})}_{h,\id} \sim  e^{-i\pi h} b^{4h} \chi_\id\left(\frac{2\pi i}{\beta_L}\right) \;G^{(\text{Schw})}_h\left(\tilde{t}_E=-i b^2 z,\tilde{\beta}=b^2\beta_L\right) \\
	G^{(\text{Schw})}_h\left(\tilde{t}_E,\tilde{\beta}\right) = \frac{1}{2\pi^2 Z^{(\text{Schw})}(\tilde{\beta})}\int d\mu(k_1) d\mu(k_2)   \frac{\prod_{\pm\pm}\Gamma(h\pm ik_1\pm ik_2)}{\Gamma(2h)} e^{- \tilde{t}_E(k_1^2-k_2^2)-\tilde{\beta}k_2^2}
\end{gather}
We have here extracted a normalizing factor of the vacuum character
\begin{gather}
	\chi_\id\left(\frac{2\pi i}{\beta_L}\right) \sim 4\pi\sqrt{2} b^3 e^{\frac{1}{24}\beta_L} Z^{(\text{Schw})}\left(\tilde{\beta}=b^2\beta_L\right),
\end{gather}
where $Z^{(\text{Schw})}$ is the Schwarzian partition function given in \eqref{eq:ZSchw}.

$G^{(\text{Schw})}_h$ is the result in \cite{Mertens:2017mtv} for the Schwarzian thermal two-point function, expressed in Euclidean time $\tilde{t}_E$, with $\Re \tilde{t}_E>0$ \footnote{The prefactor $ e^{-i\pi h} b^{4h}$ appears naturally in the Schwarzian when we rescale time and analytically continue to Lorentzian signature, since the operators transform nontrivially under scaling, but we find it more convenient for the discussion below to separate it.}. It is normalized to give a pure power law in the limit of small $\tilde{t}_E$, namely $G^{(\text{Schw})}_h\sim \tilde{t}_E^{-2h}$. Writing this in terms of $z$ and combining with the factor $e^{-i\pi h}b^{4h}$, the block $\mathcal{F}^{(\tilde{N})}_{h,\id}$ has the usual short-distance behavior $z^{-2h}$. This result is valid for $0<\Im z < \beta_L$, and the result for real $z$ (with the current operator ordering) is obtained in the limit $\Im z \to 0^+$.\footnote{In fact, if we take real $z$, by expanding $C_0$ to higher orders we find that the integral over $k_1$ is rendered convergent by an `$i \epsilon$' appearing automatically with the correct sign: the relevant correction to $\log C_0$ is proportional to $(b^2\log b)\, k_1^2$.}

Just as in \eqref{eq:sc4pt}, we can evaluate the canonical two-point function $G^{(\text{Schw})}$ in a semiclassical limit, which here means $\tilde{\beta},\tilde{t}_E \ll 1$. The calculation is much the same as led to \eqref{eq:sc4pt}, with the integral dominated by the region where both $k_1$ and $k_2$ are close to the value $\pi/\tilde{\beta}$ corresponding to the thermodynamic energy, and $k_1-k_2$ is of order one \footnote{When taking this limit we assumed that $h$ is order one. One can also consider a semiclassical limit with large $h\sim c$ which is more complicated \cite{Goel:2018ubv} but encodes some simple bulk backreaction effects.}
\begin{equation}\label{eq:SCSchw}
	G^{(\text{Schw})}_h\left(\tilde{t}_E,\tilde{\beta}\right) \sim \left(\tfrac{\tilde{\beta}}{\pi} \sin\left(\tfrac{\pi}{\tilde{\beta}}\tilde{t}_E\right) \right)^{-2h}
\end{equation} 

Finally, a nice aspect of the expression \eqref{eq:leftmovingres} (and similarly \eqref{eq:SchwMicro}) is that it interpolates between the semiclassical behavior of equation \eqref{eq:SCSchw} and the late time behavior. At late times the approximation leading to \eqref{eq:SCSchw} fails signaling that in this regime strong coupling Schwarzian effects are important. This happens in our context for times $\tilde{t}_E \gg c$. Expression \eqref{eq:leftmovingres} gives the asymptotics 
\beq
G_h^{\rm (Schw) } (\tilde{t}_E,\tilde{\beta}) \sim \tilde{t}_E^{-3}, ~~~~\tilde{t}_E \gg c
\eeq
where we omitted a time independent prefactor that depends on $c$, $\tilde{\beta}$ and $h$. In the $\tilde{\beta}\to \infty$ limit this power changes to $G_h^{\rm (Schw) } \sim \tilde{t}_E^{-3/2}$. In the context of the Schwarzian theory this was observed in \cite{Bagrets:2016cdf, Mertens:2017mtv}. In the context of 2d CFT this behavior of conformal blocks was observed numerically in \cite{Chen:2017yze} and analytically \cite{Collier:2018exn}. The advantage of \eqref{eq:leftmovingres} is then that it interpolates between different regimes.

\subsection{Time ordering and retarded two-point function}\label{sec:TORTPF}

Including both left- and right-moving pieces, we have our result for the normalized grand-canonical two-point function in the near-extremal limit:
\begin{equation}\label{eq:2ptTO}
	\frac{\langle \op(z,\bar{z})\op(0,0)\rangle_{\beta_L,\beta_R}}{Z_{\beta_L,\beta_R}} \sim e^{-i\pi h} b^{4h} \left[\frac{\beta_R}{\pi}\sinh\left(\frac{\pi}{\beta_R}\bar{z}\right)\right]^{-2\bar{h}} G^{(\text{Schw})}_h\left(-ib^2 z,b^2\beta_L\right)
\end{equation}
Note that the ordering is important here, with the insertion of $\op(z,\bar{z})$ coming after that of $\op(0,0)$, and furthermore that this is valid for $0<\bar{z}<\bar{z}_*\leq \pi$. The time-ordering appeared in our derivation through the choice of necklace channel $N$, since the simplification \eqref{eq:Nblocklimit} of the blocks occurs only with the given time ordering. Here, the other ordering differs only by a phase from swapping the operator insertions in the left-moving block\footnote{Alternatively, we could take the original result and set $(z,\bar{z})\rightarrow (-z,-\bar{z})$, so the right-moving block produces a phase $e^{2\pi i \bar{h}}$. This is equivalent for integer spin, $\bar{h}-h\in \ZZ$.} (though we will see that things get slightly trickier for out of time order correlators with more operator insertions):
\begin{equation}\label{eq:2ptOTO}
	\frac{\langle \op(0,0)\op(z,\bar{z})\rangle_{\beta_L,\beta_R}}{Z_{\beta_L,\beta_R}} \sim e^{+i\pi h} b^{4h} \left[\frac{\beta_R}{\pi}\sinh\left(\frac{\pi}{\beta_R}\bar{z}\right)\right]^{-2\bar{h}} G^{(\text{Schw})}_h\left(+ib^2 z,b^2\beta_L\right)
\end{equation}

 It will be useful to consider also the retarded correlator
\begin{equation}
	G_R(z,\bar{z}) = -i \langle [\op(z,\bar{z}),\op(0,0)] \rangle  \;\Theta\!\left(t=\tfrac{1}{2}(\bar{z}-z)\right),
\end{equation}
which for $0<\bar{z}<\bar{z}_*$ and $z= -b^{-2}\tilde{t}<0$ is given by
\begin{equation}\label{eq:GR1}
	G_R(z,\bar{z}) \sim b^{4h} \left[\frac{\beta_R}{\pi}\sinh\left(\frac{\pi}{\beta_R}\bar{z}\right)\right]^{-2\bar{h}} 2\Im\left[e^{-i\pi h} G^{(\text{Schw})}_h\left(- i b^2 z, b^2\beta_L\right)\right].
\end{equation}
Note in particular that with this definition of $G^{(\text{Schw})}_h$, the retarded correlator is not simply proportional to the retarded correlator in the Schwarzian theory, due to the additional phase in our choice of normalization. On the opposite side of the lightcone, with $-\bar{z}_*<\bar{z}<0$ and again $z= -b^{-2}\tilde{t}<0$, we have
\begin{equation}\label{eq:GR2}
	G_R(z,\bar{z}) \sim b^{4h} \left[\frac{\beta_R}{\pi}\sinh\left(-\frac{\pi}{\beta_R}\bar{z}\right)\right]^{-2\bar{h}} 2\Im\left[e^{+i\pi h} G^{(\text{Schw})}_h\left(-i b^2 z, b^2\beta_L\right)\right].
\end{equation}
In the semiclassical limit of the Schwarzian \eqref{eq:SCSchw}, the phase is precisely $e^{-i\pi h}$, so this piece vanishes to to leading order in that limit. Away from the lightcone, when $|\bar{z}|\gg \beta_R$, the correlation functions are exponentially suppressed.

We should note that the full correlator also contains lightcone singularities when $z$ is an integer multiple of $2\pi$, and our result is not strictly valid parametrically close to these lightcones. However, the strength of the singularity decays exponentially in time, and their contribution becomes negligible in the Schwarzian limit after smearing the operators by any fixed amount\footnote{We can smear either by slightly displacing the insertions in Euclidean time, or for retarded correlators by integrating against a more general smooth function, for example by taking partial waves as in the next section.}.  In the conformal block expansion, the singularities arise from an infinite sum over double-twist exchanges; our argument for dominance of the vacuum block applies in any compact region bounded away from lightcone singularities, where the sum over blocks converges uniformly.

\subsection{Partial waves}\label{sec:cftpw}

For direct comparison with a gravitational two-dimensional nAdS$_2$ dual, we should consider the partial waves, which are correlation functions of the Fourier modes of operators:
\begin{equation}\label{eq:Omodes}
	\mathcal{O}_\ell (t) = \int_0^{2\pi} \frac{d\varphi}{2\pi} ~e^{ i \ell \varphi} \mathcal{O}(z=\varphi-t,\bar{z}=\varphi+t)
\end{equation}

For this, we cannot consider time-ordered correlators, because the lightcone singularity at $\bar{z}=0$ is not integrable (at least, for $h>\frac{1}{2}$). However, the retarded correlator does not suffer from the same problem, when we interpret the singularity as a distribution. The expectation value of the commutator can be thought of as a discontinuity across a branch cut in the Euclidean time plane, and we can define the partial waves by deforming the integral to a contour passing below the cut, around the branch point, and back above it, while avoiding the singularity at the branch point itself. In practice, since this prescription for the integral is analytic, we can simply perform the integral for $h<\tfrac{1}{2}$ when it converges, and analytically continue to general $h$.

Since the typical scale on which the correlator varies in the $z$ direction is of order $c$, the integral over $\varphi$ at fixed $t$ can, to good approximation, be replaced by an integral over $\bar{z}$ at fixed $z=-t$:
\begin{align}
	G_\ell(t) &= \int_{-\pi}^{\pi} \frac{d\varphi}{2\pi} ~e^{ i \ell \varphi}G_R(z=\varphi-t,\bar{z}=\varphi+t)\\
	&\sim e^{-i\ell t}\int_{-\pi}^{\pi} \frac{d\bar{z}}{2\pi} ~e^{ i \ell \bar{z}} G_R(z=-2t,\bar{z})
\end{align}
The dominant contribution comes from close to the lightcone, with $\bar{z}$ of order $\beta_R$, both from $\bar{z}>0$ \eqref{eq:GR1} and $\bar{z}<0$ \eqref{eq:GR2}:
\begin{gather}\label{eq:partialWaves}
	G_\ell(t) \sim  e^{-i\ell t}\mathcal{N}_\ell\, 2\Im\left[G^{(\text{Schw})}_h\left(2ib^2 t ,b^2\beta_L\right)\right] \\
	\begin{aligned}
		\mathcal{N}_\ell &=\left(\frac{2\pi}{\beta_R}\right)^{2\bar{h}-1} \frac{\Gamma\left(\bar{h}+\tfrac{\ell\beta_R}{2\pi i}\right)\Gamma\left(\bar{h}-\tfrac{\ell\beta_R}{2\pi i}\right)}{\Gamma(2\bar{h})} \frac{\sin\left(\pi\bar{h}+\tfrac{1}{2}i\ell\beta_R \right)}{\sin\left(\pi\bar{h}\right)} \\
		&\sim \left(\frac{2\pi}{\beta_R}\right)^{2\bar{h}-1} \frac{\Gamma\left(\bar{h}\right)^2}{\Gamma(2\bar{h})}
	\end{aligned}
\end{gather}
We see that each individual partial wave is proportional to the retarded correlator of the Schwarzian. The approximation to $\mathcal{N}_\ell$ in the second line is valid when $\ell\beta_R\ll 1$; it is spin independent because such modes cannot resolve the details of the angular dependence, effectively seeing a delta-function on the lightcone.

To quantify the error we introduced in our approximation of the integral, we can Taylor expand the correlation function:
\begin{equation}\label{eq:KKmodescorr}
	\begin{gathered}
		G_R(z=\varphi-t,\bar{z}=\varphi+t) = G_R(\bar{z}-2t,\bar{z}) \\
		\qquad = G_R(-2t,\bar{z}) -\tfrac{1}{2}\bar{z} \partial_t G_R(-2t,\bar{z}) + \cdots
	\end{gathered}
\end{equation}
The factor of $\bar{z}$ results in an additional factor of $\beta_R$ after integrating, while the time derivative results in a factor of $c^{-1}$. The neglected corrections are therefore suppressed by a relative factor of $\beta_R/c$. We will later see that this is characteristic of interactions with graviton Kaluza-Klein modes.

\subsection{Higher-point functions and OTOC}\label{sec:generalcorr}

We now discuss some salient points for the generalization of the above results to higher point functions. This is fairly straightforward, but introduces some new ingredients: specifically, the choice of operator ordering becomes more important, and there are new kinematic regimes where an identity block need not dominate. To illustrate these new ideas, we discuss the computation of the four point function of pairwise identical operators. 

In the canonical ensemble we want to compute  
\begin{equation}
	\langle \op_A(z_1,\bar{z}_1)\op_A(z_2,\bar{z}_2)\op_B(z_3,\bar{z}_3)\op_B(z_4,\bar{z}_4) \rangle_{\beta_L,\beta_R} 
\end{equation}
for operators with dimensions $(h_A,\bar{h}_A)$ and $(h_B,\bar{h}_B)$ (considering other time orderings later). As before, we use lightcone coordinates for the locations of the operators, $z_i = \varphi_i - t_i$ and $\bar{z}_i = \varphi_i + t_i$ for $i=1,\ldots 4$. We will take all times to be large of order $t \sim c$ and choose the angles such that $0<\bar{z}<2\pi$ for all insertions.

As in section \ref{sec:rmsdvb}, we must first identify the relevant blocks in the $\beta_R\to 0$ limit by considering the right moving sector. Once again, we insert complete sets of states at circles of constant angle between every operator insertion, generalizing the dual necklace ($\widetilde{N}$) channel in equation \eqref{eq:necklaceBlock1}, here with four sets of states. We consider first a configuration with $0<\bar{z}_1<\bar{z}_2\ldots <\bar{z}_4<2\pi$ where all $\bar{z}_i$ and $\bar{z}_{ij}\equiv \bar{z}_i-\bar{z}_j$ are fixed as we take the $\beta_R\to 0$ limit. Descendants then drop out and the right moving dual necklace block is proportional to 
\begin{equation}
	\bar{\mathcal{F}}^{(\tilde{N})}_{\bar{h}_1,\ldots, \bar{h}_4} \propto  \exp\left[-\frac{2\pi}{\beta_R}\left(\bar{z}_{21} \bar{h}_1+\bar{z}_{32}\bar{h}_2+\bar{z}_{43}\bar{h}_3+(2\pi-\bar{z}_{41})\bar{h}_4\right)\right] \,.
\end{equation}
As in section \ref{sec:rmsdvb}, we find the dominant block by minimizing the kinematic combination in the exponent over values of $\bar{h}_i$ allowed by the fusion rules.

Since the right-moving block is exponentially suppressed for fixed $\bar{z}_{ij}$, these configurations will not in fact be relevant for correlation functions of partial waves. Instead, we must consider kinematic regimes where some $\bar{z}_{ij}$ scale proportionally to $\beta_R$, which we can split into three cases:
\begin{description}
\item[Case 1:] Identical operators $\op_A$ and $\op_B$ are close together in pairs, but the separation between pairs is of order one. Concretely, we could have $\bar{z}_{12}, \bar{z}_{43}$ each of order $\beta_R$, with $\bar{z}_{32}$ order one.
\item[Case 2:] Each $\op_A$ is close to one of the $\op_B$ and the pairs are order one separated. For example, we take $\bar{z}_{32},2\pi-\bar{z}_{41}$ small, with $\bar{z}_{21},\bar{z}_{43}$ order one.
\item[Case 3:] All operators are close to each other with all $\bar{z}_{ij}$ small.
\end{description}

\textbf{Case 1:} There is a unique choice of intermediate operators so the block is not exponentially suppressed in $\beta_R$, with the identity propagating when the $\bar{z}$ separation is order one. Namely, we must take $\mathcal{O}_1 =\op_A$, $\op_2=\id$, $\op_3 =\op_B$ and $\op_4=\id$. As in section \ref{sec:rmsdvb} we can write a formula for the right-moving blocks which is valid when $\bar{z}_{12}, \bar{z}_{43} \sim \beta_R$, by including descendants of $\op_1=\op_A$ and $\op_3=\op_B$:
\begin{equation}\label{eq:4ptright}
	\bar{\mathcal{F}}^{(\tilde{N})}_{\op_A, \id,\op_B,\id}(\bar{z}_{ij},\beta_R) \sim e^{\frac{(2\pi)^2}{\beta_R}\frac{c}{24}} \left[\frac{\beta_R}{\pi}\sinh\left(\frac{\pi}{\beta_R}\bar{z}_{12}\right)\right]^{-2\bar{h}_A}\left[\frac{\beta_R}{\pi}\sinh\left(\frac{\pi}{\beta_R}\bar{z}_{34}\right)\right]^{-2\bar{h}_B}
\end{equation}
This is just the product of separate vacuum two-point functions on an infinite line, along with a term encoding the Casimir energy of the vacuum. The four point function is then given by the $\widetilde{N}$ identity blocks
\begin{equation}\label{eq:case1}
\lb \op_A\op_A\op_B\op_B\rb_{\beta_L,\beta_R} \sim	\mathcal{F}^{(\tilde{N})}_{\op_A, \id,\op_B,\id}(z_{ij},\beta_L)\bar{\mathcal{F}}^{(\tilde{N})}_{\op_A, \id,\op_B,\id}(\bar{z}_{ij},\beta_R) 
\end{equation}
up to exponentially small corrections

\textbf{Case 2:} The blocks are exponentially suppressed unless the identity appears in the intermediate channels with finite $\bar{z}$ separation, namely $\mathcal{O}_1 =\id$ and $\op_3 =\id$. But the fusion rules for the identity would then simultaneously demand that $\mathcal{O}_2=\mathcal{O}_A$ and $\mathcal{O}_2=\mathcal{O}_B$ (and similarly for $\mathcal{O}_4$), which cannot both be satisfied in the same block. The leading contribution is therefore suppressed by either $e^{-\frac{2\pi}{\beta_R}\bar{z}_{21}\bar{h}_\text{gap}}$ or $e^{-\frac{2\pi}{\beta_R}\bar{z}_{43}\bar{h}_\text{gap}}$, and this region does not contribute to partial waves.

\textbf{Case 3:} When all operators are separated by $\bar{z}_{ij}$ of order $\beta_R$, there is no suppression in the $\beta_R\to 0$ limit as long as $\op_4=\id$. This means we have $\op_1=\op_A$ and $\op_3=\op_B$ (for the given the ordering of operators in $\bar{z}$), but we must include all intermediate operators in the sum over $\op_2$.  At this point, we could make an additional assumption that we have a holographic 2d CFT, where 't Hooft factorization applies, which suppresses non-vacuum channels by small OPE coefficients of order $c^{-1}$. Nonetheless, as we will see in more detail later, in partial waves the contribution from case 3 is parametrically small  regardless of such factorization, since it involves a small kinematic regime in the integral over angles.

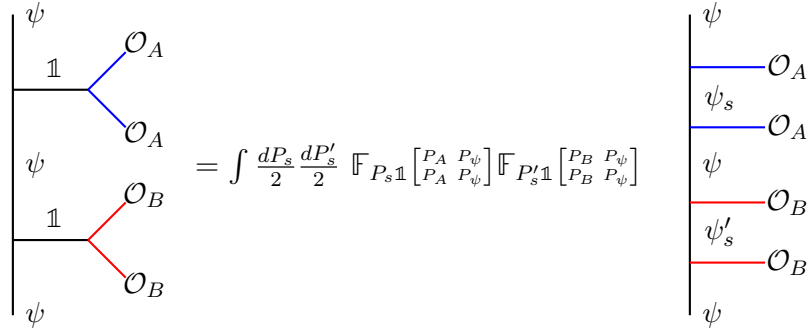
\begin{figure} [t!]
\begin{center}
\begin{tikzpicture}
\draw[thick] (0,-2) -- (0,2);
\draw[thick] (0,1) -- (1,1);
\draw[thick,blue] (1,1) -- (1.5,1.5);
\draw[thick,blue] (1,1) -- (1.5,1-0.5);
\draw[thick] (0,-1) -- (1,-1);
\draw[thick,red] (1,-1) -- (1.5,-1+0.5);
\draw[thick,red] (1,-1) -- (1.5,-1-0.5);
\draw (1.75,1.6) node  {\small$\mathcal{O}_A$};
\draw (1.75,1-0.6) node  {\small$\mathcal{O}_A$};
\draw (1.75,-1+0.6) node  {\small$\mathcal{O}_B$};
\draw (1.75,-1-0.6) node  {\small$\mathcal{O}_B$};
\draw (0.3,2) node  {\small$\psi$};
\draw (0.3,-2) node  {\small$\psi$};
\draw (0.3,0) node  {\small$\psi$};
\draw (0.55,1.3) node  {\small$\id$};
\draw (0.55,-1+.3) node  {\small$\id$};
\draw (5.4,0) node  { $=\int \frac{dP_s}{2}\frac{dP'_s}{2} ~ \fusion_{ P_s \id} \sbmatrix{P_A & P_\psi \\ P_A & P_\psi}\fusion_{ P'_s \id} \sbmatrix{P_B & P_\psi \\ P_B & P_\psi}$};
\draw[thick] (9+0,-2) -- (9+0,2);
\draw[thick,blue] (9+0,1.3) -- (9+1,1.3);
\draw[thick,blue] (9+0,0.5) -- (9+1,0.5);
\draw[thick,red] (9+0,-1.3) -- (9+1,-1.3);
\draw[thick,red] (9+0,-0.5) -- (9+1,-0.5);
\draw (9.3,0) node  {\small $\psi$};
\draw (9.4,0.89) node  {\small $\psi_s$};
\draw (9.4,-0.89) node  {\small $\psi'_s$};
\draw (9+0.3,2) node  {\small$\psi$};
\draw (9+0.3,-2) node  {\small$\psi$};
\draw (9+1.3,1.3) node  {\small$\mathcal{O}_A$};
\draw (9+1.3,0.5) node  {\small$\mathcal{O}_A$};
\draw (9+1.3,-1.3) node  {\small$\mathcal{O}_B$};
\draw (9+1.3,-0.5) node  {\small$\mathcal{O}_B$};
\end{tikzpicture}
\end{center}
\vspace{-0.5cm}
\caption{\small Diagram of the fusion transformation that was used to compute the four-point left moving vacuum block in the appropriate limit. The blue (red) line corresponds to the external $\mathcal{O}_A$ ($\mathcal{O}_B$) insertion. }
\label{fig:fusion4pt}
\end{figure}

The conclusion is that \textbf{Case 1} will give the dominant contribution to the correlators of partial waves, through the identity block \eqref{eq:case1}, with right-moving piece given by \eqref{eq:4ptright}. We next need to compute the left moving torus block by transforming back to the direct necklace channel, as in section \ref{sec:leftmovblock2pt}. The only difference is that now the fusion transformation has to be applied twice, one to each pair $\op_A\op_A$ and $\op_B\op_B$. Instead of spelling out the details, we show the relevant fusion transformation in figure \ref{fig:fusion4pt} for the case of the microcanonical ensemble calculation. The torus block is obtained similarly, with an additional integral over the state $|\psi\rb$, weighted by a Boltzmann factor and Schwarzian density of states. Taking the appropriate limits and using the expression \eqref{CMMid} we obtain
\beq
\mathcal{F}^{(\tilde{N})}_{h,\id} \sim  e^{-i\pi h_A-i\pi h_B} b^{4h_A+4h_B} \chi_\id\left(\frac{2\pi i}{\beta_L}\right) \;G^{(\text{Schw})}_{h_A,h_B}\left(\tilde{t}_i=-i b^2 z_i,\tilde{\beta}=b^2\beta_L\right) 
\eeq
where anticipating its interpretation we defined the function appearing in the right hand side as
\begin{eqnarray}
G^{(\text{Schw})}_{h_A,h_B} &=& \frac{1}{2\pi^2 Z^{(\text{Schw})}(\tilde{\beta})}\int d\mu(k_1) d\mu(k_2)d\mu(k_3)e^{-\tilde{t}_{21}k_1^2-\tilde{t}_{43}k_2^2 - (\tilde{\beta}-\tilde{t}_{21}-\tilde{t}_{43})k_3^2}\nonumber\\
&&\hspace{-1cm}   \times\frac{\prod_{\pm\pm}\Gamma(h_A\pm ik_1\pm ik_3)}{\Gamma(2h)} \frac{\prod_{\pm\pm}\Gamma(h_B\pm ik_2\pm ik_3)}{\Gamma(2h)} 
\end{eqnarray}
with $\tilde{t}_{21} \equiv -ib^2(z_2-z_1)$ and $\tilde{t}_{43} \equiv -ib^2(z_4-z_3)$. Comparing with the notation in figure \ref{fig:fusion4pt} we defined $P_s=b k_1$, $P'_s=bk_2$ and $P_\psi=b k_3$. This function that appears in the torus block is exactly the same as the Schwarzian time ordered four point function, analogously to the previous result \eqref{eq:leftmovingres}. Putting all together we obtain the full four point function in the near extremal CFT limit as 
\begin{eqnarray}
	\frac{\lb \op_A\op_A\op_B\op_B\rb_{\beta_L,\beta_R}}{Z_{\beta_L,\beta_R}} &\sim&e^{-i\pi h_A-i\pi h_B} b^{4h_A+4h_B}G^{(\text{Schw})}_{h_A,h_B}\left(\tilde{t}_i=-i b^2 z_i,\tilde{\beta}=b^2\beta_L\right)  \nonumber\\
	&&\hspace{-2.5cm}\times \left[\frac{\beta_R}{\pi}\sinh\left(\frac{\pi}{\beta_R}\bar{z}_{12}\right)\right]^{-2\bar{h}_A}\left[\frac{\beta_R}{\pi}\sinh\left(\frac{\pi}{\beta_R}\bar{z}_{34}\right)\right]^{-2\bar{h}_B}
\end{eqnarray}
for a choice of $\bar{z}_{1,2,3,4}$ that falls under Case 1 above. Note that the choice of operator ordering is important, as we will discuss more in a moment.

From the diagrammatic rules defined in \cite{Mertens:2017mtv} to compute Schwarzian correlators, and the fact that the fusion transformation of the block is always done in pairs, it is clear that this connection between Virasoro blocks and the Schwarzian theory will generalize to higher point correlators.

To summarize, the important configurations have identical operators in pairs, separated in $\bar{z}$ by order $\beta_R$ (Case 1), but with finite separation between each pair. In such a case, the correlator is dominated by an appropriate identity block, which gives the Schwarzian correlator in the left-moving sector, times a product of cylinder two-point functions in the right-moving sector. The vacuum dominance can fail when we do not have nearby pairs of identical operators (such as Case 2), when the correlator is in any case exponentially suppressed. It can also fail when two or more pairs of identical operators come within a $\bar{z}$ separation of order $\beta_R$ (Case 3), unless we make the additional assumption of factorization.

\subsubsection*{Partial waves} 

As anticipated above, the situation improves when we integrate over angles (which is equivalent to leading order in $\beta_R$ to integrating over $\bar{z}$). When computing partial waves correlators as in section \ref{sec:cftpw} we can fix the position of one insertion ($\bar{z}_1=0$ for example) and integrate over the remaining coordinates. In Case 1 the dominant contribution comes from fixing two other coordinates, $\bar{z}_{21}$ and $\bar{z}_{43}$ to accuracy $\beta_R$. In Case 3 the dominant contribution comes from all coordinates being of order $\beta_R$. Therefore the angular integral will produce, schematically, factors of 
\beq
\int_{\rm Case~1} \prod_{i=1}^4 d\bar{z}_i (\ldots) \sim \beta_R^2 (\ldots),~~~~{\rm vs.}~~~~\int_{\rm Case~3} \prod_{i=1}^4 d\bar{z}_i (\ldots) \sim \beta_R^3 (\ldots),
\eeq
where the additional factor of $\beta_R$ simply comes from the small region of $\bar{z}_{32}$ for which vacuum dominance fails. This implies that any non vacuum block, that due to the $\beta_R \to 0$ limit can only contribute for configurations of type 3, will come with an extra factor of $\beta_R$. Therefore we conclude that the partial wave correlators are dominated only by the vacuum blocks that produce the Schwarzian answer, namely 
\beq\label{eq:pw4ptres}
	\lb \op^{\ell_A}_A(t_1)\op^{\ell_A}_A(t_2)\op^{\ell_B}_B(t_3)\op^{\ell_B}_B(t_4)\rb_{\beta_L,\beta_R} \propto \mathcal{N}_{\ell_A}\mathcal{N}_{\ell_B}\, G^{(\text{Schw})}_{h_A,h_B}\left(2ib^2 t_i ,b^2\beta_L\right) ,
\eeq
where we have taken a `retarded' combination of time-orderings for each pair of operators to make sense of the partial wave integral. The prefactors $\mathcal{N}_\ell$ are given by \eqref{eq:partialWaves} for each pair of operators. 
 
 There are several types of corrections to our formula \eqref{eq:pw4ptres}. The first are corrections to the vacuum block itself, arising from the angular integral as explained around equation \eqref{eq:KKmodescorr}, and (new for higher-point functions) also from ignoring right-moving descendants of the vacuum, most relevant for `Case 3' configurations, where four operators have $\bar{z}$ separation of order $\beta_R$. Both sources of error result in corrections of order $\beta_R/c$, and admit a bulk interpretation as exchanges of graviton KK modes. In addition, we have corrections from non-vacuum blocks, most importantly from `Case 3' configurations, which are of order $\beta_R$ times the relevant OPE coefficients. In a gravitational description, these arise from interactions (including, but not limited to, interactions with matter KK modes), as will be explained in section \ref{sec:dimredmatKK}. In a truly holographic theory with a weakly interacting bulk dual, the OPE coefficients multiplying the non-vacuum blocks are small, giving an additional suppression.

\subsubsection*{Out-of-time-order correlators (OTOC)}

For the arguments above, it was important that we were considering a time-ordered four-point function. The simplification of left-moving `direct' necklace channel blocks generalizing \eqref{eq:Nblocklimit} is only valid when the order of operators in the choice of channel corresponds to the correct time ordering. On the other hand, the dominance and simplification of the vacuum block in the right-moving sector relies on choosing the $\widetilde{\text{Necklace}}$ channel corresponding to `spatial' ordering, meaning that we place operators in order of $\bar{z}$ (though this is not so important for the nearby pairs of identical operators, since we have included the relevant intermediate descendants to give the $\sinh$ in \eqref{eq:4ptright}, for example). We already saw a version of this in the difference between \eqref{eq:2ptTO} and \eqref{eq:2ptOTO}, though the effect of this was simple to account for since we needed only exchange the order of nearby identical operators.

For the time-ordered four-point function in the configurations such as those (Case 1) which dominate the partial wave correlators, we can always arrange for the ordering in time to match the ordering in $\bar{z}$ (up to exchanging the pairs of nearby identical operators). However, this is no longer true for out-of time ordered (OTOC) orderings such as
\begin{equation}
	\langle \op_A(z_1,\bar{z}_1)\op_B(z_3,\bar{z}_3)\op_A(z_2,\bar{z}_2)\op_B(z_4,\bar{z}_4) \rangle_{\beta_L,\beta_R},
\end{equation}
when pairs of identical operators do not appear consecutively. We need to add an additional `braiding' move to exchange operator order before inserting the simplified necklace block.

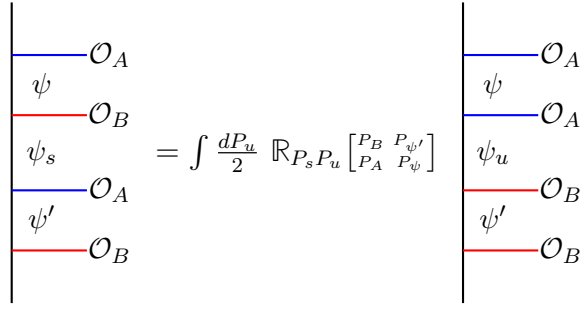
\begin{figure} [t!]
\begin{center}
\begin{tikzpicture}
\draw[thick] (+0,-2) -- (+0,2);
\draw[thick,blue] (+0,1.3) -- (+1,1.3);
\draw[thick,red] (+0,0.5) -- (+1,0.5);
\draw[thick,red] (+0,-1.3) -- (+1,-1.3);
\draw[thick,blue] (+0,-0.5) -- (+1,-0.5);
\draw (0.4,0) node  {\small $\psi_s$};
\draw (0.4,0.89) node  {\small $\psi$};
\draw (0.4,-0.89) node  {\small $\psi'$};
\draw (+1.3,1.3) node  {\small$\mathcal{O}_A$};
\draw (+1.3,0.5) node  {\small$\mathcal{O}_B$};
\draw (+1.3,-1.3) node  {\small$\mathcal{O}_B$};
\draw (+1.3,-0.5) node  {\small$\mathcal{O}_A$};
\draw (3.8,0) node  { $=\int \frac{dP_u}{2} ~ \braid_{ P_s P_u} \sbmatrix{P_B & P_{\psi'} \\ P_A & P_\psi}$};
\draw[thick] (6+0,-2) -- (6+0,2);
\draw[thick,blue] (6+0,1.3) -- (6+1,1.3);
\draw[thick,blue] (6+0,0.5) -- (6+1,0.5);
\draw[thick,red] (6+0,-1.3) -- (6+1,-1.3);
\draw[thick,red] (6+0,-0.5) -- (6+1,-0.5);
\draw (6.4,0) node  {\small $\psi_u$};
\draw (6.4,0.89) node  {\small $\psi$};
\draw (6.4,-0.89) node  {\small $\psi'$};
\draw (6+1.3,1.3) node  {\small$\mathcal{O}_A$};
\draw (6+1.3,0.5) node  {\small$\mathcal{O}_A$};
\draw (6+1.3,-1.3) node  {\small$\mathcal{O}_B$};
\draw (6+1.3,-0.5) node  {\small$\mathcal{O}_B$};
\end{tikzpicture}
\end{center}
\vspace{-0.5cm}
\caption{\small Diagram of the braiding transformation that is need to compute out-of-time-ordered correlators. }
\label{fig:braid}
\end{figure}

The argument in the right moving sector is unchanged for the OTOC, so we need only consider the left-moving vacuum block $\mathcal{F}^{(\tilde{N})}_{\op_A, \id,\op_B,\id}(z,\beta_L)$, where the $\widetilde{N}$ channel arranges the operators in $\bar{z}$ ordering: $AABB$. First, we apply the same sequence of moves as before to rewrite the $\widetilde{N}$ block as a direct necklace channel block. However, this block inherits the $AABB$ ordering, but the necklace block only simplifies if the channel matches the $ABAB$ time ordering. The change of order of two of the operators 
is achieved using the `R-matrix' as shown in figure \ref{fig:braid}. After braiding, the new necklace channel block with $ABAB$ ordering becomes a scaling block, and we have new $z$ dependence coming from integrating over the internal index in the R-matrix. This is a building block that we can now use for any higher point OTOC, constructing any time ordering with repeated applications. The R-matrix is simply related to the fusion matrix and therefore we can use again the formula derived by Ponsot and Teschner \cite{Ponsot:1999uf}. 

The limit of the R-matrix required in the Schwarzian limit was computed in Appendix B of \cite{Mertens:2017mtv}. In the notation of figure \ref{fig:braid} we take all intermediate states to be near extremal $P_\psi = b k$, $P_{\psi'}=b k'$, $P_{s/u}=b k_{s/u}$ and the external operators $h_A$ and $h_B$ to be light. Then the R-matrix becomes  
\beq\label{rmatrix}
\braid_{P_s P_u} \sbmatrix{P_B & P_{\psi'} \\ P_A & P_\psi}  \! =\! \frac{\rho_0(bk_u)}{2\pi b^3}  \left| \frac{ \Gamma(h_A+ ik \pm i k_u) \Gamma(h_B +i k'\pm i k_u)}{\Gamma(h_A+ ik' \pm i k_s) \Gamma(h_B +i k\pm i k_s)}\right|\; \sixj{h_A}{h_B}{k'}{k}{k_s}{k_u}
\eeq
where we defined
\begin{equation}
	\begin{aligned}
		&\sixj{h_A}{h_B}{k'}{k}{k_s}{k_u} = \mathbb{W}(k_s, k_u ; h_A + i k,h_A - i k, h_B - i k',h_B + i k') \\
		 &\qquad\times\,\sqrt{\Gamma(h_A \pm i k' \pm ik_s)\Gamma(h_B \pm i k \pm ik_s)\Gamma(h_B \pm ik'\pm ik_u)\Gamma(h_A \pm i k \pm i k_u)}.
	\end{aligned}	
\end{equation}

This object that we obtain as a limit of the Ponsot-Teschner general formula is the 6j-symbol of the classical $\mathfrak{sl}(2)$ computed by Groenevelt \cite{groenevelt}. $ \mathbb{W}(\alpha,\beta;a,b,c,d)$ is the Wilson function also defined by Groenevelt \footnote{The definition we are using is 
\beq
{\tiny \mathbb{W}(\alpha,\beta;a,b,c,d) \equiv \frac{\Gamma(d-a)~_4F_3\Big[\mbox{\small$\begin{array}{cccc}\! a+i\beta \!\! & \!\! a-i\beta \!\! &\!\! \tilde{a}+i \alpha &\!\! \tilde{a}-i\alpha \!\hspace{-2pt} \\[-1mm] \!\!\!\! a+b \!\! &\!\!\!\!\! a+c \!\! &\!\! 1+a-d \!\hspace{-2pt} \end{array}$} ; 1 \Big]}{\Gamma(a+b)\Gamma(a+c)\Gamma(d\pm i \beta) \Gamma(\tilde{d}\pm i\alpha)}  + (a\leftrightarrow d),}
\eeq
where $\tilde{d}=(b+c+d-a)/2$ and $\tilde{a}=(a+b+c-d)/2$.}. The prefactor of $(2\pi b^3)^{-1}$ combines with the factor of $b^2$ in \eqref{eq:rho0limit} and the factor of $b$ in the measure $dP_u= b dk_u$ to ensure that the braiding  does not change the factors of $b$, so the OTOC is of the same order as the TOC.

When this transformation is applied to the left moving block, the final answer is proportional to a function $G^{OTOC}_{h_A,h_B}\left(\tilde{t}_i=-i b^2 z_i,\tilde{\beta}=b^2\beta_L\right)$ that reproduces again the Schwarzian OTOC. We will not write it down here but the final expression can be found in equation (1.28) of \cite{Mertens:2017mtv}.\footnote{The appearance of the 6j-symbol in Schwarzian OTOC found in \cite{Mertens:2017mtv} was reproduced using the BF formalism in \cite{Blommaert:2018oro, Iliesiu:2019xuh} and using the boundary particle approach in \cite{Suh:2019uec}. Moreover, it was verified in \cite{Lam:2018pvp} that the semiclassical limit of this kernel gives the Dray-'t Hooft shockwave S-matrix, reproducing the semiclassical calculation of the OTOC of \cite{Maldacena:2016upp}.} 

The full OTOC at fixed angles (with a configuration such as Case 1 where the identity block dominates, and spatial ordering is $AABB$) including left- and right-moving blocks is given by
\beq
\frac{\lb \op_A\op_B\op_A\op_B\rb_{\beta_L,\beta_R}}{Z_{\beta_L,\beta_R}} \propto  G^{OTOC}_{h_A,h_B}\left(\tilde{t}_i=-i b^2 z_i,\tilde{\beta}=b^2\beta_L\right)  \bar{\mathcal{F}}^{(\tilde{N})}_{\op_A, \id,\op_B,\id}(\bar{z}_i).
\eeq
We can use this formula to obtain correlators of partial waves by integrating over angles and the result is that the OTOC for the CFT is proportional to the Schwarzian OTOC, similarly to all previous cases. We can also use this for the OTOC without integrating over angles, where the two insertions of $\op_A$ are at the same spacetime location (up to small shifts to regulate), and likewise the two insertions of $\op_B$.

We would like to stress that this is the first derivation of OTOC in 2d CFT where the vacuum dominance approximation is justified, thanks to the small parameter $\beta_R$. This is in contrast to \cite{Roberts:2014ifa}, for example, for which there is no clear justification for vacuum dominance \cite{Chang:2018nzm}. This shows in particular that the gravitational S-matrix of 3d gravity coupled to matter is exactly given in the near extremal limit by the 6j-symbol of $\mathfrak{sl}(2)$. This can be thought of as a controlled derivation, in a certain limit, of the universal gravitational scattering proposed in \cite{Jackson:2014nla}.  

As a final application we can compute the semiclassical limit of the (out-of-time-ordered) left moving Virasoro block using the braiding matrix \eqref{rmatrix}. The block is proportional to the Schwarzian correlator and the semiclassical limit of the exact OTOC was derived in \cite{Lam:2018pvp} giving 
\begin{eqnarray}
\frac{\mathcal{F}^{OTOC}_{\op_A, \id,\op_B,\id}(z,\beta_L)}{\mathcal{F}^{TO}_{\op_A, \id,\op_B,\id}(z,\beta_L)} &\sim& \eta^{-2h_A}U(2h_A,1+2h_A-2h_B,\eta^{-1}), \\
\eta&\equiv& \frac{i\tilde{\beta}}{2\pi}\frac{e^{-i\frac{\pi}{\tilde{\beta}} (\tilde{t}_3+\tilde{t}_4-\tilde{t}_1-\tilde{t}_2)}}{\sin \frac{\pi \tilde{t}_{12}}{\tilde{\beta}}\sin \frac{\pi \tilde{t}_{34}}{\tilde{\beta}}}
\end{eqnarray}
with $\tilde{t}=-ib^2 t$ and $\tilde{\beta}=b^2 \beta$ and $U(a,b,c)$ being the confluent hypergeometric function. The semiclassical limit corresponds to late times with $\eta$ small (but larger than other small parameters). The analog for the time ordered four point block gives simply a product of \eqref{eq:SCSchw} two point functions. This matches with the semiclassical calculation done in \cite{Maldacena:2016upp} and our derivation of this expression from a conformal block explains the origin of equation (4.14) and (4.17) of \cite{Chen:2016cms}. When this formula is expanded at small $\eta$ it gives the maximal $\lambda_L=\frac{2\pi}{\beta}$ Lyapunov exponent saturating the chaos bound of \cite{Maldacena:2015waa} \footnote{The fixed angle OTOC grows with time exponentially with rate $\lambda_L=2\pi/ \beta_L\approx\pi/\beta$ while the s-wave correlator grows with rate $\lambda_L=2\pi/\beta$. This is consistent with the bounds derived in \cite{Mezei:2019dfv} (see section 5).}.

Finally, since we have control over the OTOC calculation (thanks to the $\beta_R\to0$ limit), we could also attempt to compute the quantum Lyapunov spectrum \cite{Gharibyan:2018fax} which is given by OTOC between four arbitrary operators. This is related to inelastic scattering in the bulk and an analogous chaos bound applies \cite{Turiaci:2019nwa}. We can compute OTOC between different operators in the near extremal limit, which picks the intermediate channel with lower twist. The bound implies their spin $s$ has to satisfy $s\leq 2$ and also puts a bound on their OPE coefficients, but we leave a detailed analysis for future work.

\section{Near extremal rotating BTZ black holes}\label{sec:BTZ}

\subsection{The BTZ black hole and its AdS$_2$ throat}\label{secbtzandthroat}

We begin our analysis of the dual gravitational physics with a look at the classical BTZ black hole and its thermodynamics. This is a solution to Einstein gravity in three dimensions with negative cosmological constant $\Lambda=-\ell_3^{-2}$ (the subscript on $\ell_3$ distinguishing it from the two-dimensional AdS length encountered later), with metric
\begin{gather}\label{eq:BTZmetric}
	ds^2 = - f(r) dt^2 + \ell_3^2\frac{dr^2}{f(r)} + r^2 \Big(d\varphi - \frac{r_-r_+}{r^2} dt\Big)^2, \\
	f(r)=\frac{(r^2-r_+^2)(r^2-r_-^2)}{r^2}.
\end{gather}
We are using a dimensionless time coordinate $t$ such that the asymptotic metric is proportional to $-dt^2+d\varphi^2$, and $f$ has dimensions of length squared. The inner and outer horizons at $r=r_\pm$, with $0<r_-<r_+$ (for $J>0$), are related to the energy and angular momentum of the black hole by
\begin{equation}
	M = \frac{r_+^2+r_-^2}{8 G_N \ell_3}, \qquad J = \frac{r_+r_-}{4G_N\ell_3}.
\end{equation}
The mass here is the dimensionless energy conjugate to the time $t$, with zero energy defined such that empty AdS$_3$ has $M=-\frac{\ell_3}{8G_N}$, corresponding to the Casimir energy of the dual.

Using the Brown-Henneaux relation $c= \frac{3\ell_3}{2G_N}$ for the central charge of the dual CFT in the classical limit, and including one-loop corrections to the energy from the Casimir energy of gravitons \cite{Maxfield:2019hdt}, the horizons can be related very simply to the CFT parameters introduced in \eqref{eq:parcft} and \eqref{eq:parcft2}:
\begin{equation}
 	P = Q \frac{r_+-r_-}{2\ell_3},\qquad 	\bar{P} = Q \frac{r_+ + r_-}{2\ell_3}
\end{equation}

Applying the standard gravitational thermodynamics, we find the classical black hole temperature (from surface gravity), angular potential (from horizon angular velocity) and entropy (from the area):
\begin{equation}
	T =\beta^{-1} = \frac{r_+^2-r_-^2}{2\pi \ell_3 r_+}, \quad \Omega = \frac{r_-}{r_+}, \quad S = \frac{2\pi r_+}{4G_N}
\end{equation}
The angular potential is the chemical potential for angular momentum, related to the parameter $\theta$ used in section \ref{sec:inv} by  $\theta = i \beta\Omega$. Writing this in terms of left- and right-moving temperatures $\beta_L=(1+\Omega)\beta$, $\beta_R=(1-\Omega)\beta$, we have
\begin{equation}
	\beta_L = \frac{2\pi \ell_3}{r_+-r_-}, \quad \beta_R = \frac{2\pi \ell_3}{r_++r_-}.
\end{equation}

The BTZ solution we have given is smooth outside an event horizon as long as $r_\pm$ are real, which imposes the extremality bound $M\geq |J|$. We are interested in near-extremal black holes, close to saturating this bound, so the difference between $r_+$ and $r_-$ goes to zero.
 This is a low temperature limit, in which the thermodynamic quantities approach the following:
\begin{equation}\label{nethermo4}
	M- J \sim \frac{\pi^2\ell_3}{8G_N}T^2,\qquad S\sim \pi \sqrt{\frac{J\ell_3}{G_N}} + \frac{\ell_3\pi^2}{4G_N} T
\end{equation}
This can be compared with the thermodynamics of the Schwarzian theory, for which the energy and entropy are given by
\begin{equation}\label{thermoSchw}
	E-E_0 = 2 \pi^2 C T^2, \qquad S=S_0 + 4 \pi^2 CT.
\end{equation}
Defining the central charge using the Brown-Henneaux relation $c= \frac{3\ell_3}{2G_N}$, we recover the values $C=\frac{c}{24}$, $E_0 =  J$ and  $S_0 = 2 \pi \sqrt{\frac{c}{6}J}$ found in \eqref{eq:NEPF} of section \ref{sec:inv}.

If we simply take a near-extremal limit of \eqref{eq:BTZmetric}, fixing $r_+$ and taking $r_-\to r_+$ with fixed $r$, we find the extremal BTZ metric:
\begin{equation}\label{eq:extrBTZ}
	ds^2 \sim - \frac{(r^2-r_+^2)^2}{r^2} dt^2 + \ell_3^2 \frac{r^2}{(r^2-r_+^2)^2}dr^2 + r^2 \left(d\varphi - \tfrac{r_+^2}{r^2} dt\right)^2
\end{equation}
However, in this limit we obscure the most physically important and interesting region of the spacetime. We expect a large class of black holes in a near-extremal limit to develop a near-horizon AdS$_2$ throat region \cite{Almheiri:2016fws, Nayak:2018qej, Moitra:2019bub, Castro:2018ffi}, and this is no exception. To zoom in on this throat, we must scale the radius with the temperature, introducing a new coordinate $\rho$ via
\begin{equation}
	r=\tfrac{1}{2}(r_++r_-) + \tfrac{1}{2}(r_+-r_-)\rho,\quad T\sim \frac{r_+-r_-}{\pi\ell_3}\to 0,
\end{equation}
and fix $\rho$ as we go to low temperature. We then find the geometry
\begin{equation}\label{eq:NearHorizonMetric}
	ds^2 \sim \frac{\ell_3^2}{4}\left[-(\rho^2-1) (2\pi T)^2 dt^2 +\frac{d\rho^2}{(\rho^2-1)}\right]+r_+^2 \left(d\varphi-dt + \tfrac{\ell_3}{r_+}\rho \,\pi T\, dt \right)^2,
\end{equation}
which is a fibration of a circle over AdS$_2$ of radius $\ell_2 = \frac{1}{2}\ell_3$, as found in \cite{Achucarro:1993fd}. This itself is a solution of pure 3d gravity, sometimes called the `self-dual orbifold' \cite{Coussaert:1994tu,Balasubramanian:2009bg}, and is the BTZ analog of the near horizon extremal Kerr geometry \cite{Bardeen:1999px}. The metric has an enhanced isometry group $SL(2,\RR)\times U(1)$. In this near-horizon region, the effects of finite temperature are still visible, and for temperatures of order $c^{-1}$ quantum effects become of leading importance, as we now briefly review.

If we ignore corrections to the AdS$_2$ throat geometry, an infinite-dimensional symmetry emerges, of diffeomorphisms relating the boundary of AdS$_2$ to the physical time, which is spontaneously broken to $SL(2,\RR)$ by the choice of AdS$_2$ vacuum. We therefore expect the low-energy physics to be described by a Goldstone mode parameterizing $\operatorname{Diff}(S^1)/SL(2,\RR)$. 

The simplest option would be for the theory of the Goldstone mode to preserve the $SL(2,\RR)$ and $\operatorname{Diff}(S^1)/SL(2,\RR)$ symmetries, but no such theory exists. It was shown in \cite{Jensen:2011su} it is not possible to have a quantum mechanical system with an exact $SL(2,\RR)$ symmetry. Accordingly, in \cite{Maldacena:1998uz} it is shown that gravity in AdS$_2$ cannot support finite energy excitations (see also \cite{Almheiri:2014cka}). We are therefore forced to include the leading order explicit breaking of this symmetry; the resulting theory of the pseudo-Goldstone is precisely the Schwarzian theory \cite{Maldacena:2016upp}. For low temperatures $T\sim c^{-1}$ (the `gap temperature' of \cite{Preskill:1991tb} at which the thermodynamic description of near extremal black holes was thought to break down), this mode becomes strongly coupled, so quantum effects are of leading importance.

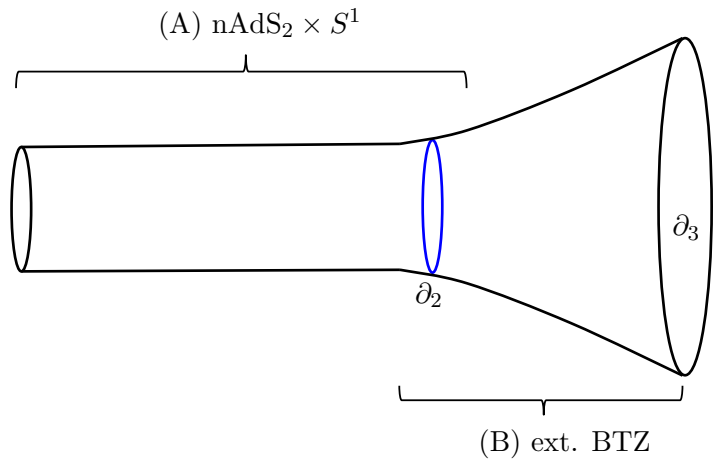
\begin{figure}
\centering
\begin{tikzpicture}[scale=0.9]
\pgftext{\includegraphics[scale=0.5]{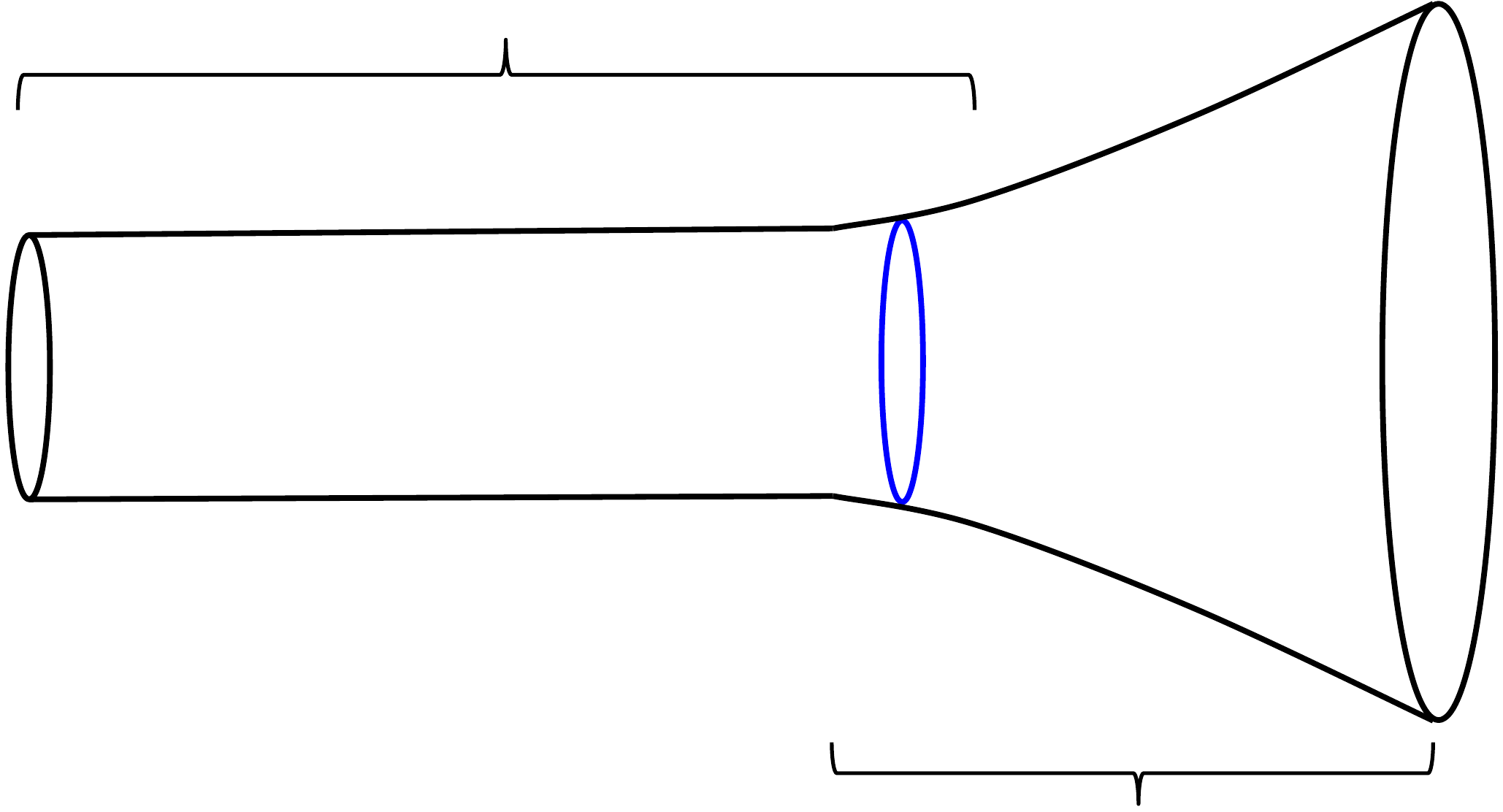}} at (0,0);
\draw (-1.5,3) node  {\small (A) nAdS$_2\times S^1$};
\draw (3,-3.2) node  {\small (B) ext. BTZ};
\draw (4.8,0) node  {\small $\partial_3$ };
\draw (1,-1) node  {\small $\partial_2$ };
\end{tikzpicture}
\caption{\small The near extremal BTZ geometry at fixed time, from the horizon (leftmost circle) to the asymptotically AdS$_3$ boundary (rightmost circle $\partial_3$). In region (A) the geometry is approximately nearly AdS$_2\times S^1$ (described by JT gravity), and in (B) the geometry is approximately extremal BTZ and the physics is classical. In the overlap between (A) and (B), we introduce the boundary $\partial_2$ (the blue line) where the Schwarzian mode lives.\label{fig:nebtzregions}}
\end{figure}

This Schwarzian theory will give a good description of the physics in the region where corrections to the AdS$_2$ fibration \eqref{eq:NearHorizonMetric} are small. This is true as long as $r-r_+ \ll r_+$, corresponding to $\rho \ll \frac{r_+}{\ell_3 T}$, which is the region (A) illustrated in figure \ref{fig:nebtzregions}. In another region, labeled by (B) in figure \ref{fig:nebtzregions}, quantum fluctuations are suppressed, and we can accurately describe the physics classically, on a fixed background, which is approximately extremal BTZ. This is the region $\rho\gg 1$, corresponding to $r-r_+ \gg \ell_3 T$. The two regions (A), where JT gravity is useful, and (B), where the geometry is classical, have a parametrically large region of overlap. Somewhere in this region, we can place an artificial boundary $\partial_2$ of the AdS$_2$ region. The physics inside this boundary will be described by JT gravity, which induces a Schwarzian theory living on $\partial_2$. This will be matched to the asymptotic boundary $\partial_3$ of AdS$_3$, where the dual CFT lives, by the classical physics between $\partial_2$ and $\partial_3$.

\subsection{A two-dimensional theory}

To describe the dynamics in the AdS$_2$ throat described above, we first analyze the two-dimensional theory arising on dimensional reduction of three-dimensional gravity. 

\subsubsection{Dimensional reduction} \label{SecDimRed}

Our `parent' theory is simply three-dimensional Einstein-Hilbert gravity, with action (here in Euclidean signature)
\begin{equation}
	I_\text{EH} = -\frac{1}{16 \pi G_N} \left[ \int_M d^3 x \,\sqrt{g_3} \Big(R_3+\tfrac{2}{\ell_3^2}\Big) + 2  \int_{\partial M}d^2x \sqrt{\gamma} \Big(\kappa_3-\tfrac{1}{\ell_3}\Big) \right],
\end{equation}
where the metric $g_3$, scalar curvature $R_3$ and boundary extrinsic curvature $\kappa_3$ carry subscripts to distinguish them from the two-dimensional quantities introduced presently. The boundary conditions are those standard in AdS/CFT, and we have included the Gibbons-Hawking term and counterterm ($\gamma$ is the induced metric on $\partial M$) to render the action finite.

We will study configurations of this theory with a $U(1)$ symmetry, with Killing field $\partial_\varphi$. We use coordinates $x^a$ and $\varphi$, where $a$ runs over two indices (which will be identified with the time and radial coordinates in the BTZ configuration), and the coordinate $\varphi$ in the symmetry direction is periodically identified as $\varphi\sim \varphi+2\pi$. With this restriction, a general metric can be written in Kaluza-Klein form
\begin{equation}\label{eq:KKmetric}
	g_3 = g_2 + \Phi^2(d\varphi+A)^2,
\end{equation}
where the two-dimensional metric $g_2$, gauge field one-form $A$ and scalar dilaton $\Phi$ are independent of $\varphi$, depending only on the two-dimensional coordinates $x^a$.

With this ansatz, the three-dimensional quantities are written in terms of two-dimensional fields ($\Box_2$ is the Laplacian corresponding to $g_2$) as
\begin{equation}
\begin{aligned}
	R_3 &= R_2 - \tfrac{1}{4}\Phi^2 F_{ab}F^{ab}-2\Phi^{-1} \Box_2 \Phi \\
	\sqrt{g_3} &= \Phi \sqrt{g_2} \\
	d^2x\sqrt{\gamma} &= \Phi dt_E d\varphi \\
	\kappa_3 &= \kappa_2+ \Phi^{-1}\partial_n\Phi.
\end{aligned}
\end{equation}
For the boundary terms, we also assume that the location of the cutoff $\partial M$ is independent of $\varphi$.
The equation for the measure on the boundary $d^2x\sqrt{\gamma}$ is then simply a definition of the proper time coordinate $t_E$ parameterizing the boundary of the two-dimensional manifold. $\partial_n$ denotes the unit normal derivative at the boundary.

Inserting this in the three-dimensional action and integrating over $\varphi$, we find the two-dimensional Einstein-Maxwell-dilaton action:
\begin{equation}
	I_\text{EH} = -\frac{2\pi}{16\pi G_3}\left[\int d^2 x\sqrt{g_2}\Phi\Big(R_2 - \tfrac{1}{4}\Phi^2 F_{ab}F^{ab}+\tfrac{2}{\ell_3^2}\Big) +2\int_\partial ds\Phi(\kappa_2-\tfrac{1}{\ell_3}) \right]
\end{equation}
Here, $F=dA$ is the field strength of the Kaluza-Klein gauge field, and all indices are contracted with $g_2$. Note in particular that a total derivative term $\Box_2\Phi$ exactly cancels with the extra term in the Gibbons-Hawking boundary action. The three-dimensional diffeomorphisms become two-dimensional diffeomorphisms, as well as gauge transformations arising from $x^a$-dependent translations in the $\varphi$ direction, so that $A$ is a compact $U(1)$ gauge field.

It remains only to discuss the boundary conditions of the reduced theory. The three-dimensional boundary conditions are that the induced metric $\gamma$ is proportional (with holographic renormalisation parameter $\epsilon$) to a chosen metric on which the CFT dual lives; we may always choose a flat metric $ds^2 = dt_E d\varphi$, so the boundary condition is $\gamma = \epsilon^{-2} dt_E d\varphi$. Reducing to two dimensions, this implies that the dilaton is just the holographic renormalization constant, $\Phi = \epsilon^{-1}$. We can think of this condition as defining the location of the cutoff surface, from which we subsequently determine $t_E$. The metric boundary condition then gives the period of $t_E$ in terms of the inverse temperature $\beta$; the proper length of the boundary is $L_\partial = \epsilon^{-1}\beta$.

The gauge field is a little more subtle. At first sight, it looks like we should impose $A=0$ on the boundary, but in fact we can only do this locally. The boundary value of $A$ acts as a background gauge field for the one-dimensional dual quantum mechanics, which by gauge transformations we can always set to be trivial locally; however, this is not true globally when the boundary is a circle, since our gauge transformation must be single-valued around the circle. This leaves a single piece of gauge invariant data, the holonomy $\theta = \int A$. This is an angle, defined up to shifts by $2\pi$, since we can take gauge transformations that wind an integer number of times around $U(1)$ as we go round the circle. Tracing the holonomy back to its higher-dimensional origin, we find that it is precisely the twist angle we encountered in section \ref{sec:inv}, from the periodicity condition $(t_E,\varphi)\sim (t_E,\varphi+2\pi)\sim (t_E+\beta,\varphi+\theta)$. In summary, the reduction of the standard AdS/CFT boundary conditions becomes the `Dirichlet boundary conditions',
\begin{equation}\label{eqdbc}
	\text{Dirichlet BC:}\qquad \left.\Phi\right|_\partial =\epsilon^{-1}, \quad L_\partial = \epsilon^{-1}\beta, \quad \int_\partial A=\theta \,,
\end{equation}
imposed on the asymptotically AdS$_3$ boundary.

\subsubsection{Integrating out the gauge field}\label{SecU1Red}

We can simplify this theory still further, due to the simplicity of gauge fields in two dimensions, in particular the absence of a propagating photon. Since our gauge field is abelian, its action is quadratic, so we could simply integrate it out; however, we will nonetheless find it convenient to first rewrite the Maxwell action in a first-order $BF$ formalism\footnote{This approach becomes extremely useful for near extremal black holes in higher dimensions with non-abelian gauge fields \cite{higherdim}\cite{Maldacena:2019cbz}.}, often used for the solution of nonabelian gauge theories in two dimensions \cite{Blau:1991mp, Witten:1991we}. 
\begin{gather}
	I_\text{Maxwell} = \frac{1}{32G_3}\int d^2x \sqrt{g_2} \Phi^3 F_{ab}F^{ab} \rightarrow  \int \left[i\mathcal{J} F +\tfrac{1}{2} \mu \mathcal{J}^2\right] \\
	\mu = 8G_3d^2x \sqrt{g_2} \Phi^{-3}
\end{gather}
We have here introduced the auxiliary scalar field $\mathcal{J}$, and the Maxwell theory now manifestly depends on the metric and dilaton only through the measure two-form $\mu$.  Integrating out the auxiliary field, we find that it is related to the field strength by
\begin{equation}\label{eq:calJ}
	F= i \mathcal{J}\mu,
\end{equation}
and using this to substitute for $\mathcal{J}$ we return to the original second-order Maxwell action. Note that for real Euclidean geometries, $A$ is real and $\mathcal{J}$ is imaginary, but this is reversed for solutions with a continuation to real Lorentzian geometries.

Now, we may instead integrate out the gauge field. We find that it imposes the constraint that $\mathcal{J}$ is a constant, which by examining the expansion of the gauge field near the boundary we can identify as the charge in the dual CFT. The three-dimensional origin of this charge is the ADM angular momentum $J$. The first term in the action then is a total derivative, which becomes the holonomy on the boundary via $\int F = \theta$, so we find
\begin{equation}
	\mathcal{J} =J,\quad  I_\text{Maxwell} = i J \theta + \tfrac{1}{2}J^2\int \mu.
\end{equation}
The last step is to sum over spins $J$, which are forced to be integers by the periodicity of $\theta$. The result is a partition function of the Maxwell piece:
\begin{equation}\label{eq:ZMaxwell}
	Z_\text{Maxwell} = \sum_{J=-\infty}^\infty e^{-i J \theta - \tfrac{1}{2}J^2\int \mu }
\end{equation}
As an alternative way to see this result, we can quantize the theory `radially', and fix a static gauge; we find that the theory becomes the quantum mechanics of a free particle propagating on a circle parameterized by the holonomy $\theta$, and $J$ labels the momentum modes, which are eigenstates of the radial Hamiltonian with energy proportional to $J^2$. We then compute $Z_\text{Maxwell}$ by preparing the ground state with $J=0$ at the origin, evolving for Euclidean time proportional to $\int \mu$, and evaluating the wavefunction at fixed $\theta$. See \cite{Picken:1988ev, Mertens:2018fds} for more details, and the nonabelian generalization of a particle propagating on a group manifold.\footnote{By Poisson resummation, we can also express the result \eqref{eq:ZMaxwell} as a sum over winding numbers for the particle propagating on a circle, in which different terms are related by $\theta\to \theta + 2n\pi$. Different values of the index $n$ correspond to different three-dimensional topologies, which are the same set of $SL(2,\ZZ)$ black holes mentioned in footnote \ref{sl2zFootnote}.} 

From this result, we see that each term in the sum over $J$ has the very nice property that it contributes a local effective action for the dilaton and metric. Since the sum over terms does not retain this property, it is most natural to perform the path integral not with fixed holonomy `Dirichlet' boundary conditions, but with fixed angular momentum `Neumann' boundary conditions. This is just the change of ensemble in the partition function we saw already in section \ref{sec:inv}, where we pick out a Fourier mode of $\theta$ by an integral $\int d\theta e^{iJ\theta}Z_\text{Maxwell}$. From the bulk point of view, we can describe this by adding a local boundary counterterm to the action,
\begin{equation}
	I \longrightarrow I + iJ\int_\partial \theta
\end{equation}
and using fixed $J$ boundary conditions; this counterterm is precisely what is required to make the variational problem with the new boundary conditions well-defined. Integrating out the Maxwell field then gives a local effective action, giving rise to an Einstein-dilaton theory:
\begin{gather}\label{eq:AOaction}
	I_J = -\frac{2\pi}{16\pi G_N}\left[\int d^2 x\sqrt{g_2}\left( \Phi R_2 -U(\Phi) \right) +2\int_\partial ds\,\Phi (\kappa_2-\tfrac{1}{\ell_3}) \right] \\
	U(\Phi) = \tfrac{1}{2}(8G_N J)^2 \Phi^{-3}-\tfrac{2}{\ell_3^2}\Phi
\end{gather}

For calculations at fixed holonomy, corresponding to the grand canonical ensemble with fixed chemical potential, we can now simply compute at fixed spin $J$ with this local effective action, before summing over $J$. Here we assumed the absence of matter charged under the gauge field, which are inevitably present from Kaluza-Klein modes breaking the $U(1)$ symmetry. In the presence of such fields, the gauge field can play a more nontrivial role.

The action in equation \eqref{eq:AOaction} was originally derived by Achucarro and Ortiz \cite{Achucarro:1993fd}, but we hope that our presentation clarifies the role of the $U(1)$ gauge field and its boundary condition \footnote{See also \cite{Almheiri:2016fws} and \cite{Gaikwad:2018dfc} for a different approach and \cite{Iliesiu:2019lfc} for a recent more thorough analysis of JT gravity coupled to 2d Yang Mills theory.}.

\subsubsection{The near-extremal limit and the Schwarzian theory}\label{SecNELST}

The description we have given so far did not require a near-extremal limit. Our next step is to take such a limit, and extract the dynamics of the near-AdS$_2$ region from the action given in \eqref{eq:AOaction}, which is in the class of models studied by \cite{Almheiri:2014cka}.

We first look for solutions where the metric $g_2$ is exactly AdS$_2$. This requires $\Phi$ to be a constant $\Phi=\Phi_0$ at a zero of the potential $U(\Phi_0)=0$, and the AdS$_2$ radius $\ell_2$ is determined by $U'(\Phi_0)$:
\begin{gather}
	U(\Phi_0)=0 \implies \Phi_0 = \sqrt{4G_N \ell_3 J} \\
	U'(\Phi_0) = -\frac{2}{\ell_2^2} \implies \ell_2 = \tfrac{1}{2}\ell_3
\end{gather}
Reinstating the gauge field using
\begin{equation}\label{eq:AdS2F}
	F = 8 i J G_N \Phi^{-3}  d^2x\sqrt{g_2} = \frac{i}{\sqrt{\ell_3^3J G_N}} \mathrm{vol}_{\text{AdS}_2},
\end{equation}
where $\mathrm{vol}_{\text{AdS}_2}$ is the volume form (area element) of AdS$_2$, and Wick rotating to Lorentzian signature, we find that this is precisely the `self-dual orbifold' geometry \eqref{eq:NearHorizonMetric} we found in the near-horizon of near-extremal rotating BTZ.

To incorporate the leading order fluctuations away from AdS$_2$, we expand the dilaton around its extremal value, writing
\begin{equation}
	\Phi = \Phi_0 + 4G_N\phi,
\end{equation}
and find the terms in the action  linear in $\phi$:
\begin{gather}
	I_J = -S_0 \chi + I_\text{JT}[g_2,\phi] + \cdots \label{eq:NEaction}\\
	\chi = \frac{1}{4\pi}\int d^2x \sqrt{g_2} R_2 + \frac{1}{2\pi} \int_{\partial_2} \kappa_2, \quad S_0 = 2\pi\frac{\Phi_0}{4G_N} = 2\pi \sqrt{\frac{\ell_3 J}{4G_N}}\\
	I_\text{JT} = - \frac{1}{2}\int d^2 x \sqrt{g_2}\, \phi\left(R_2 +\frac{2}{\ell_2^2}\right)-\int_{\partial_2} \phi(\kappa_2-\tfrac{1}{\ell_2}) \label{eq:JTaction}
\end{gather}
The leading order term is the two-dimensional Einstein-Hilbert action, which is topological, proportional to the Euler characteristic $\chi$ of spacetime. The next term is the action for Jackiw-Teitelboim (JT) gravity. Subsequent terms are suppressed in the limit $\phi\ll S_0$.

In writing the boundary terms here, we have implicitly introduced a new boundary, denoted $\partial_2$ (indicated by the blue line in figure \ref{fig:nebtzregions}), because the JT approximation is only valid deep in the near-horizon region where $\Phi$ is close to $\Phi_0$. We choose to place this new, artificial boundary $\partial_2$ at a curve of constant dilaton $\phi=\phi_\partial$, where $1\ll \phi_\partial \ll S_0$. We have introduced boundary terms in the JT action so that this boundary condition is a good variational problem, with an intrinsic counterterm so that the action has a finite limit when we take $\phi_\partial\to \infty$; these boundary terms are not physical, so we must add equal and opposite terms to the action for the exterior of the throat. For the metric, we would like a boundary condition that fixes the proper length $L_2$ of $\partial_2$, but this is not freely chosen; rather, it is determined by solving the theory classically outside the AdS$_2$ throat region, between $\partial_2$ and $\partial_3$ where the physical boundary conditions are imposed.

\subsubsection{Solving the theory outside the throat}\label{SecNELST2}

When we are far enough from the black hole so that $\phi$ is no longer much smaller than $S_0$, we must keep the full dilaton potential, but can solve the theory classically; we can think of $\phi^{-1}$ as a running coupling, so quantum fluctuations are large in the throat region where $\phi$ is of order one, but small outside the artificial boundary $\partial_2$ where $\phi\gg 1$. To good approximation, we can therefore simply solve the equations of motion of the theory with appropriate boundary conditions at the boundary of AdS$_3$ $\partial_3$, find the boundary conditions induced at $\partial_2$, and finally evaluate the on-shell action between $\partial_2$ and $\partial_3$.

The solution to our two-dimensional Einstein-dilaton theory \eqref{eq:AOaction} is simply the dimensional reduction of the extremal BTZ metric \eqref{eq:extrBTZ}. In the coordinates of section \ref{secbtzandthroat}, the dilaton is given by the radial coordinate $\Phi =r$, with $\Phi_0 = r_+$, and the boundary $\partial_2$ is in the region $\ell_3 T \ll r-r_+ \ll r_+$.

First, we can read off the boundary conditions at $\partial_2$ from the metric \eqref{eq:extrBTZ}, by evaluating the proper length of the Euclidean time circle $L_2=\frac{r^2-r_+^2}{r}\beta\sim 2(r-r_+)\beta$ at fixed $r = r_+ + 4G_N \phi_\partial$. The physical content of this boundary condition, independent of our choice of $\phi_\partial$,  is the ratio of the length $L_2$ to the boundary dilaton:
\begin{equation}\label{eq:AdS2BC}
	\frac{L_2}{\phi_\partial} = 8G_N\beta = \frac{24}{c}\ell_2\beta
\end{equation}

Our second task is to evaluate the on-shell action between $\partial_2$ and $\partial_3$, where we include boundary terms at $\partial_2$ to cancel the terms we added to the JT action \eqref{eq:JTaction}. We can compute this from the extremal BTZ solution with $\partial_2$ located at constant radius, even though the configurations are perturbations of this classical solution. Outside the throat, we can treat the deviations to linear order, since higher orders do not contribute finite action. Because we are perturbing around a classical solution, the linearized deviation of the bulk action is a total derivative. But the boundary actions at $\partial_3$ and at $\partial_2$ (to cancel the boundary terms in the Euler characteristic and JT actions) have been chosen to make the variational problem well-posed, which means that the variation of the boundary action precisely cancels the variation of the bulk action after integrating by parts.

 The terms contributing finite action in the limit of interest are the bulk action between $\partial_2$ and $\partial_3$, the boundary term at $\partial_3$, and the boundary term at $\partial_2$ from the Euler characteristic topological action:
\begin{align}
		-\frac{2\pi}{16\pi G_N}\int_{\partial_2}^\infty d^2 x\sqrt{g_2}\left( \Phi R_2 -U(\Phi) \right) &\sim 2  J \beta - 8 \sqrt{\ell_3^{-1}G_N J}\beta \phi_\partial + \cdots \nonumber \\
		-\frac{1}{4 G_N}\int_{\partial_3} ds\, \Phi (\kappa_2-\tfrac{1}{\ell_3}) &\sim -  J \beta \\
		-\frac{S_0}{2\pi} \int_{\partial_2} \kappa_2 &\sim 8 \sqrt{\ell_3^{-1}G_N J}\beta \phi_\partial \nonumber
\end{align}
Adding these up, we find a total action from the outside of
\begin{equation}
	I_\text{outside} \sim  J \beta,
\end{equation}
which acts only to shift the zero of energy to the BTZ extremality bound. This is a concrete example of a general result derived in Appendix A of \cite{Moitra:2019bub}.

\subsubsection{The Schwarzian theory} \label{SecNEST3}

In the previous section we argued that the dynamics of BTZ can be reduced to JT gravity living in the throat. JT gravity is a very simple theory and can be completely reduced to a boundary mode \cite{Almheiri:2014cka} with the Schwarzian action \cite{Jensen:2016pah, Maldacena:2016upp, Engelsoy:2016xyb}. 

 To see this, we can first integrate out the dilaton, which acts as a Lagrange multiplier imposing $R_2 = -\frac{2}{\ell_2^2}$, so the metric is locally AdS$_2$, which we can write in Poincar\'e coordinates $(u,z)$ as
 \begin{equation}
 	ds^2 = \ell_2^2\frac{du^2+dz^2}{z^2}.
 \end{equation}
 The bulk action vanishes when the constraint $R_2 = -\frac{2}{\ell_2^2}$ is imposed, leaving only the boundary term. This nonetheless leaves nontrivial dynamics, arising from the location of the boundary, which is determined by the reparameterization $f$ relating the coordinate $u$ to the physical time $t_E$, as $u = \tan \frac{\pi}{\beta}f(t_E)$. This is the pseudo-Goldstone mode referred to earlier, taking values in the coset $f \in \operatorname{Diff}(S^1)/SL(2,\RR)$, where the quotient removes physically equivalent configurations obtained by the isometries of AdS$_2$. The Euclidean time $t_E$ is proportional to the proper length along $\partial_2$, up to a factor of $L_2/\beta$ relating the coordinate periodicity $\beta$ with the proper length $L_2$ of the curve. Since the length of the boundary is large $L_2\gg \ell_2$, we can choose it to lie at small $z$, and obtain the extrinsic curvature in that approximation \cite{Maldacena:2016upp}:
 \begin{equation}
	\kappa_2 - \ell_2^{-1} \sim \ell_2 \frac{\beta^2}{L_2^2} \Big\{ \tan \frac{\pi}{\beta} f(t_E),t_E\Big\}
\end{equation}
Integrating this, we recover the Schwarzian action
\beq
I_\text{JT} = -C  \int_0^{\beta} dt_E \Big\{ \tan \frac{\pi}{\beta} f(t_E), t_E\Big\},
\eeq
with coefficient $C$ given by
\begin{equation}
C  = \ell_2 \beta \frac{\phi_\partial}{L_2}   = \frac{c}{24},
\end{equation}
where we finally made use of \eqref{eq:AdS2BC}, and the result is independent of the arbitrary choice of $L_2$ determining the location of the cutoff surface $\partial_2$.

When JT gravity is coupled to free matter the boundary correlators are simply given by Schwarzian expectation values of the Schwarzian bilocal as shown in \cite{Maldacena:2016upp}. This expectation value is the quantity computed in \cite{Mertens:2017mtv}. The origin of the free matter approximation will be explained below in the next section. 

Finally, we note that, while we have obtained JT gravity from a near-extremal limit of pure 3D gravity, this is not quite the end of the story once we consider nonperturbative corrections (though these will not be relevant for this work). Additional topologies, beyond those visible in pure JT gravity considered in \cite{Saad:2019lba}, become relevant, and are particularly important at very low temperature. This is the subject of work in progress \cite{nonpert}.

\subsection{Matter, correlation functions and Kaluza-Klein modes}\label{sec:sourcesmatch}

To study correlation functions in the BTZ background for comparison to the results of section \ref{sec:CFT}, we now consider the effect of adding matter. 

\subsubsection{The classical limit}

First, we study the classical limit, where the temperature is high enough ($\beta\ll c$) that backreaction is unimportant, so we are simply studying correlation functions in a fixed BTZ background. Since this geometry is a quotient of pure AdS$_3$, we can use the method of images to construct the two-point function of a free field from the corresponding result on the plane, which is determined by conformal symmetry. In the lightcone coordinates $(z,\bar{z})$, where we may choose $0<\bar{z}<2\pi$, the retarded correlator $G_R$ (for $z<0$, required so that $G_R$ is nonzero) is given by
\begin{equation}
	G_R(z,\bar{z}) = -2\sin(2\pi h) \sum_{n=0}^{\lfloor -z/2\pi\rfloor} \left(\frac{\beta_L}{\pi}\sinh\left(\frac{\pi}{\beta_L}(-z-2n\pi)\right)\right)^{-2h} \left(\frac{\beta_R}{\pi}\sinh\left(\frac{\pi}{\beta_R}(\bar{z}+2n\pi)\right)\right)^{-2\bar{h}}.
\end{equation}
The image sum over $n$ runs over the finite number of images lying in the future lightcone $\bar{z}>0,z<0$ of the origin. The $\sin(2\pi h)$ appearing in the prefactor does not break the left-right symmetry because the spin $\ell=\bar{h}-h$ is an integer, so it may also be written as $\sin(2\pi \bar{h})$ or $(-1)^\ell\sin(\pi \Delta)$.

For small $\beta_R$ (and $\bar{h}>0$), the $n>0$ terms in the image sum are exponentially suppressed relative to the $n=0$ term. This dominant term precisely reproduces the retarded correlator \eqref{eq:GR1}, where we evaluate the Schwarzian correlator in the semiclassical limit \eqref{eq:SCSchw}. As already mentioned, the result \eqref{eq:GR2}, which would usually dominate for $\bar{z}$ close to $2\pi$, is zero in the semiclassical limit.

If we were to extract the Fourier modes of this result, we would of course reproduce the partial waves \eqref{eq:partialWaves}, which are given by a normalization factor times the Schwarzian correlation function.

\subsubsection{Dimensional reduction of matter}\label{sec:dimredmatKK}

To explain the results in a more general context, we perform a dimensional reduction of matter fields. Suppose our three-dimensional theory contains a massive scalar field $\chi$, with action
\begin{equation}
I_\chi= -\frac{1}{2} \int d^3x \sqrt{g_3} \left[g_3^{\mu\nu}\partial_\mu\chi \partial_\nu\chi + V(\chi)\right]
\end{equation}
for some potential $V(\chi)=m^2\chi^2 + (\text{interactions})$ (where we have ignored boundary terms). This is dual to a CFT operator $\op$ with dimensions $h=\bar{h}=\frac{\Delta}{2}$, with $m^2\ell_3^2=\Delta(\Delta-2)$.

Given our Kaluza-Klein form \eqref{eq:KKmetric} for the three dimensional metric $g_3$, we can write the inverse metric in terms of two-dimensional fields as follows:
\begin{equation}
	g_3^{\mu\nu} = \begin{pmatrix}
		g_2^{ab} & -g_2^{ab}A_b \\
		-g_2^{ab}A_b & \Phi^{-2} + A_a g_2^{ab} A_b
	\end{pmatrix}
\end{equation}
We write the the matter field $\chi$ in terms of two-dimensional Kaluza-Klein modes, mirroring the mode decomposition \eqref{eq:Omodes} of the operator $\op$:
\begin{equation}
\chi(t,r,\varphi) = \sum_{l} e^{-i l \varphi} \chi_l (t,r)
\end{equation}
The three-dimensional action for $\chi$ then becomes an action for the complex scalars $\chi_l$ (with $\chi_{-l}=\chi_l^*$), which have charge $l$ under the Kaluza-Klein gauge field $A$:
\begin{equation}
	I_\chi = -\frac{1}{2}\int d^2x\sqrt{g_2} \;2\pi\Phi \left[\sum_l \left(g_2^{ab}D_a\chi_l D_b \chi_{-l} + (m^2+l^2\Phi^{-2}) \chi_l\chi_{-l}\right)  + \text{interactions}\right]
\end{equation}
The covariant derivative is $D_a = \partial_a-A_a\partial_\varphi = \partial_a+i l A_a$, and the interactions involve products of three or more $\chi_l$ fields with total charge adding to zero.

Deep in the AdS$_2$ throat, we can replace the dilaton $\Phi$ by the constant value $\Phi_0=r_+$, so the Kaluza-Klein modes have mass $m_l^2 = m^2 + l^2\Phi_0^{-2}$ in the AdS$_2$ region. For $\Phi_0$ of order $\ell_3$, these shifts of the mass are important for finite values of $l$, as are interactions if the field $\chi$ was originally strongly interacting in AdS$_3$. However, in the limit of very large black holes $\beta_R\ll 1$ studied in section \ref{sec:CFT}, so $\Phi_0$ is the largest parameter, we have many simplifications. For any fixed $l$, the effective mass $m_l^2$ of $\chi_l$ in the AdS$_2$ region becomes close to the original mass $m^2$. The charge of such Kaluza-Klein modes is also negligible, because the electric field \eqref{eq:AdS2F} in the AdS$_2$ region is small at very large $J$, of order $(J G_N)^{-1/2}$ in AdS units. Most importantly, the action is multiplied by an overall factor $\Phi_0$, which suppresses all interactions. For example, a cubic vertex $\lambda \chi^3$ in the potential is suppressed by a factor $\Phi_0^{-1/2}\lambda$ in the AdS$_2$ region. The same result should also hold for interactions with Kaluza-Klein modes of the graviton, so their corrections are suppressed by a factor of $\frac{G_N}{\Phi_0}$, or $\frac{\beta_R}{c}$ in the CFT variables of section \ref{sec:CFT}.

As a result, to leading order in the large $\Phi_0$ limit, we can treat the Kaluza-Klein modes as independent free fields of equal mass, and neglect their charge. Note that this is entirely different from the usual situations in Kaluza-Klein compactifications, where we can ignore the KK modes because they are heavy; here, they are instead light, but decoupled.

The parametric suppression of interactions found from this gravitational calculations reproduces the corrections to partial wave correlation functions found in section \ref{sec:CFT}. First, the corrections from graviton Kaluza-Klein modes arise from the slight smearing of the Schwarzian time when we integrate over $\varphi$, as explained at the end of section \ref{sec:cftpw}. Our CFT calculations give us a specific prediction of the full dependence on $t$ and $\varphi$, which should arise from these metric KK modes. For higher-point functions, we also have corrections from interactions of bulk fields. As explained more fully in section \ref{sec:generalcorr}, these are important only in the limited regime of kinematics when four operators (identical in pairs) approach within $\beta_R$ in the $\bar{z}$ coordinate, so are suppressed by a factor of $\beta_R$ times the coupling when we integrate over angles to compute partial waves. This is the same parametric suppression deduced from the gravitational argument.

In conclusion, to describe matter in the AdS$_2$ region, to leading order in the limit of interest we can use correlation functions of free matter coupled to JT gravity. The only ingredient left is to compute the effective `IR' conformal dimension $\Delta_S$ in this region, determining the dimension of the dual operator appearing in the Schwarzian correlation function. For this, we only need to know the relationship $\ell_3 = 2\ell_2$ between three- and two-dimensional AdS lengths:
\begin{equation}
	m^2\ell_3^2 = \Delta(\Delta-2),\quad m^2 \ell_2^2 = \Delta_S(\Delta_S-1)\implies\Delta_S = \frac{\Delta}{2} = h
\end{equation}
We have written $\Delta_S$ in terms of the left-moving dimension $h$, which is the result we expect to generalise if we were to consider matter with spin.

\subsubsection{Interpolating boundary conditions from AdS$_3$ to AdS$_2$} \label{Secmatching}

The dimensional reduction demonstrates that we can treat matter in the AdS$_2$ throat as free fields coupled to JT gravity. To compare with the 2d CFT results, we now need to propagate these correlators from the boundary of the throat to the boundary of the AdS$_3$ spacetime. We can neglect all interactions (including backreaction) in this region, so it suffices to study matter wave equations on a fixed background.

From the usual AdS/CFT dictionary, CFT$_2$ correlators can be read off from the ratio between normalizable and non-normalizable modes of the field at the asymptotic boundary $\partial_3$ (at least for the correlators we consider with pairs of identical operators). The Schwarzian correlators are similarly related to the modes at the boundary of the AdS$_2$ region $\partial_2$. The wave equation between $\partial_3$ and $\partial_2$ relates these modes, providing a map from the Schwarzian correlators to the CFT$_2$ correlators. Here, we will solve this mapping between $\partial_2$ and $\partial_3$ in frequency space, and show that for the relevant frequencies ($\omega$ of order $c^{-1}$) the map is trivial, giving a rescaling independent of $\omega$. This means that the Schwarzian correlators are directly imprinted on the asymptotic boundary of AdS$_3$. Physically, this occurs because the time to propagate from the edge of the AdS$_2$ throat to the asymptotic boundary is only of order $\ell_3$, which is very short compared to the characteristic timescale $c$ of the interactions in the deep AdS$_2$ region.

We consider a scalar field $\chi$ of mass $m^2$, and write $\Delta_\pm$ for the two roots of $\Delta(\Delta-2)=\ell_3^2 m^2$, so $\Delta_+$ is the dimension of the dual CFT operator, and $\Delta_-=d-\Delta_+$. The asymptotic expansion of $\chi$ (in frequency space, so $\chi$ is a function of the radial coordinate $r$ times $e^{-i\omega t-i\ell\varphi}$) is
\begin{equation}\label{eq:AdS3expansion}
 \chi |_{\partial AdS_3} = A_\ell(\omega) \left(\frac{r}{\ell_3}\right)^{-\Delta_-} + B_\ell(\omega) \left(\frac{r}{\ell_3}\right)^{-\Delta_+} +\cdots,
 \end{equation}
 and we can read off the correlation function from the ratio of $B$ and $A$ coefficients  (in more detail, this could be achieved by deriving an effective action, as done in a similar context in \cite{Nayak:2018qej}). For example, for a two-point function we have
 \begin{equation} \label{eq:AdS3asympt}
 	G_\ell(\omega) = \frac{\pi}{\Delta-1}\frac{B_\ell(\omega)}{A_\ell(\omega)} \,,
 \end{equation}
where the prefactor comes from applying the normalization standard in CFT$_2$, rather than the `natural' normalization in AdS, which comes from taking AdS propagators with unit strength delta-function source to the boundary. Similarly, in the asymptotically AdS$_2$ region the field will behave as 
\begin{equation}\label{eq:AdS2expansion}
 \chi |_{\partial AdS_2} = \tilde{A}_\ell(\omega) \left(\frac{4}{\ell_3}(r-r_+)\right)^{-\frac{\Delta_-}{2}} + \tilde{B}_\ell(\omega) \left(\frac{4}{\ell_3}(r-r_+)\right)^{-\frac{\Delta_+}{2}} + \cdots,
 \end{equation}
 where we have chosen the radial coordinate to give a canonical AdS$_2$ metric in the region $r_+-r_-\ll r-r_+ \ll r$. This means that for two-point functions we have
 \begin{equation} \label{eq:AdS2asympt}
 	G^{\text{AdS}_2}_\ell(\omega) = \frac{\sqrt{\pi}\Gamma\left(\tfrac{\Delta-1}{2}\right)}{\Gamma\left(\tfrac{\Delta}{2}\right)} \frac{\tilde{B}_\ell(\omega)}{\tilde{A}_\ell(\omega)}.
 \end{equation}

By solving the wave equation in the fixed BTZ background, we can express $A,B$ in terms of $\tilde{A},\tilde{B}$. The details are given in Appendix \ref{App:matter}, and we find the trivial rescaling
\beq\label{eq:boun}
\tilde{A}_{\ell}(\omega) =  \Big(\frac{2}{\pi}\beta_R\Big)^{1-\frac{\Delta}{2}} A_\ell(\omega) ,\qquad \tilde{B}_\ell(\omega) =  \Big(\frac{2}{\pi}\beta_R\Big)^{\frac{\Delta}{2}} B_\ell(\omega)
\eeq
This neither mixes normalizable and non-normalizable modes, nor adds any new frequency dependence. The result is a simple $\omega$- and $\ell$-independent rescaling factor between AdS$_3$ and AdS$_2$ correlators.

Putting this together with the ratio of the normalizing factors in \eqref{eq:AdS3asympt} and \eqref{eq:AdS2asympt}, we find
\begin{equation}
	G_\ell(\omega) \sim  \left(\frac{2\pi}{\beta_R}\right)^{\Delta-1} \frac{\Gamma\left(\tfrac{\Delta}{2}\right)^2}{\Gamma(\Delta)} 2^{-\Delta} G^{\text{AdS}_2}_\ell(\omega).
\end{equation}
This reproduces precisely the normalizing factor we found in \eqref{eq:partialWaves}, excepting the $2^{-\Delta}$, which arises from the factor of $2$ between the `lightcone' time $-z$ used to define the Schwarzian correlators and the asymptotic time $t$ used here.

\paragraph{Acknowledgements} 

We thank M.\ Berkooz, L.\ Iliesiu, E.\ Martinec, T.\ Mertens, M.\ Mezei, I.\ Papadimitriou, G.\ Sarosi and S.\ Trivedi for useful discussions. This research was supported in part by the Heising-Simons Foundation, the Simons Foundation, and National Science Foundation Grant No. NSF PHY-1748958. AG is also supported by a Dean's Competitive Fellowship for the College of Arts and Sciences at the University of Kentucky. HM is funded by a Len DeBenedictis Postdoctoral Fellowship and under NSF grant PHY1801805, and receives additional support from the University of California. GJT is supported by a Fundamental Physics Fellowship. GJT would like to thank Weizmann institute and KIAS for hospitality while part of this work was being done.

\appendix

\section{Connection with Liouville} \label{app:connliouville}
In sections \ref{sec:inv} and \ref{sec:CFT} we show that when a state of large charge (in our case angular momentum) and low temperature is considered, correlators (and the partition function itself) are dominated by vacuum blocks. Using explicit formulas for fusion matrix elements we computed the low temperature moving component of these blocks and got the Schwarzian correlators on the nose. A natural question to ask then is why did this work?

To simplify the problem we can focus on the two point function. We can rewrite the relation \eqref{CMMid} in a suggestive way in terms of Liouville theory quantities. It is straightforward after some algebra to show that the fusion matrix element is equal to 
\beq\label{idfus}
  \fusion_{ P_s \id} \sbmatrix{P_1 & P_2 \\ P_1 & P_2}  =\frac{1}{2\pi} C(\alpha_1,\alpha_2,\alpha_s) \frac{\Psi_{\rm ZZ}(0) \Psi_{\rm ZZ}(\alpha_s^*)}{\Psi_{\rm ZZ}(\alpha_1) \Psi_{\rm ZZ}(\alpha_2)}
   \eeq
where the right hand side is written in terms of $\alpha_i = \frac{Q}{2}+i P_i$. The first factor is the DOZZ formula \cite{Dorn:1994xn, Zamolodchikov:1995aa} with the usual normalization 
\beq\label{app:DOZZ}
C(\alpha_1,\alpha_2,\alpha_s) =\frac{(\pi \mu \gamma(b^2)b^{2-2b^2})^{\frac{Q-\alpha_{1+2+s}}{b}} \Upsilon(2\alpha_1)\Upsilon(2\alpha_2)\Upsilon(2\alpha_s)}{\Upsilon(\alpha_{1+2+s}-Q)\Upsilon(\alpha_{1+2-s})\Upsilon(\alpha_{1-2+s})\Upsilon(-\alpha_{1-2-s})}.
\eeq
where to shorten the notation we used $\alpha_{i\pm j\pm k} = \alpha_i \pm \alpha_j \pm \alpha_k$, and the second is the ZZ-brane boundary state wavefunction  
\beq\label{app:ZZ}
\Psi_{ZZ}\Big(\alpha=\frac{Q}{2}+iP\Big) = \frac{2^{-1/4}2 \pi ~2 i P (\pi \mu \gamma(b^2))^{-i P/b}}{\Gamma(1-2ibP)\Gamma(1-2iP/b)}
\eeq
derived in \cite{Zamolodchikov:2001ah} from a modular bootstrap analysis. The left hand side of \eqref{idfus} is a Virasoro kinematical quantity (independent of the theory) while the right hand side is written in terms of objects appearing in a specific theory, Liouville. As a check that this relation makes sense, one can see that the only theory dependent quantity, the Liouville cosmological constant $\mu$, cancels completely in the right hand side. This formula was proposed by analyzing the conformal bootstrap of Liouville theory with boundaries in \cite{Zamolodchikov:2001ah} (see their equations 6.4 and 6.5). Starting from the exact expression of the fusion matrix, the identity \eqref{idfus} was derived by Teschner and Vartanov in Appendix D.2 of \cite{Vartanov:2013ima}. This has also an interesting interpretation in the context of AGT \cite{LeFloch:2017lbt}. 

Using \eqref{idfus} it is easy to see that the torus vacuum block for the two point function used in \eqref{tvblm} is equal to the one-point function of a primary operator of Liouville theory, living in an annulus between ZZ-brane boundary states. This is required by the consistency of the boundary CFT bootstrap of \cite{Zamolodchikov:2001ah}. In the limit considered in this paper, Liouville theory between ZZ-branes is equivalent to the Schwarzian theory and Liouville primary operators are equivalent to inserting a Schwarzian bilocal field. This fact was used in \cite{Mertens:2017mtv} to compute Schwarzian correlators using 2d Liouville CFT techniques. We will leave the details of this relation for a future work, but this gives the underlying reason for the match between the results of section \ref{sec:CFT} and the Schwarzian theory. 

This is also consistent with the fact that Liouville between ZZ-branes in equivalent to the Alekseev-Shatashvili coadjoint orbit action studied in \cite{Turiaci:2016cvo, Cotler:2018zff} which in the semiclassical limit reduces to the Schwarzian action. 

\begin{center}
\textit{Schwarzian Limit}
\end{center}

The Schwarzian limit of the fusion matrix given in the form \eqref{idfus} was done in \cite{Mertens:2017mtv}. The relevant limit of the operator momentum is  $P_1= bk_1$, $\alpha_2 = b h$ and $P_s = b k_s$. The important relation here is $\Upsilon(b x) = \frac{b^{\frac{1}{2}-x}}{\Gamma(x)} f(b)$, and $\Upsilon'(0) = b^{-1/2} f(b)$,  with $f(b)$ some known function that will cancel in the end. Then it is easy to see that 
  $$
 C(\alpha_1,\alpha_2,\alpha_s)  \sim \frac{1}{b} \frac{\prod_{\pm\pm}\Gamma(h \pm ik_1 \pm i k_s)}{\Gamma(- 2 i k_1)\Gamma(2 h )\Gamma(-2 i k_s)},~~~\frac{\Psi_{\rm ZZ}(0) \Psi_{\rm ZZ}(\alpha_s^*)}{\Psi_{\rm ZZ}(\alpha_1) \Psi_{\rm ZZ}(\alpha_2)} \sim b^{4 h} \frac{\Gamma(-2 i k_1)}{\Gamma(2 i k_s)}
  $$
which gives the same expression for the fusion matrix element \eqref{eq:C0limit}. This approximation for the DOZZ formula can also be obtained by using the minisuperspace approximation of Liouville theory without having to know the full expression involving special functions.

 \section{Details on matching boundary conditions} \label{App:matter}
 
 In this appendix, we solve the wave equation for a free massive scalar field in the fixed BTZ background and work out the proportionality factors between the sources on the $AdS_3$ boundary and the asymptotic $AdS_2$ boundary, as stated in \ref{eq:boun}.

We can carry out a mode decomposition of the scalar field as
\begin{equation}
\chi (t,r,\varphi) = \sum_\ell \int_{-\infty}^\infty d\omega \ e^{-i \omega t - i\ell \varphi } \chi_{\ell,\omega}(r),
\end {equation}
The frequency space wave equation, written in terms of a new variable $z = \frac{r^2 -r_+^2}{r^2-r_-^2}$, for near-extremal BTZ written in coordinates \eqref{eq:BTZmetric} is the following:
\beq\label{eq:3dwave} 
\left[z(1-z) \frac{d^2}{dz^2} +(1-z) \frac{d}{dz}+ \frac{1}{4}\left( \frac{(\omega r_+ - \ell r_- )^2 \ell_3^2 }{z (r_+^2-r_-^2)^2} -\frac{(\omega r_- - \ell r_+ )^2 \ell_3^2 }{(r_+^2-r_-^2)^2}-\frac{\ell_3^2 m^2}{1-z} \right)\right] \chi_{\ell, \omega}(z) =0,
\eeq
See \cite{Birmingham:2001pj} for more details. In the following we will take the limit of small $T \sim r_+ - r_-$ together with the low frequency limit $\omega \sim T$, which simplifies the equation considerably. Moreover if we focus on the region between the nAdS$_2$ and AdS$_3$ boundaries we are in the regime $r-r_+\gg r_+-r_-$, for which
 \beq\label{AppBlimz}
 1-z \sim \frac{2 r_+ (r_+-r_-)}{r^2-r_+^2}. 
 \eeq
 and from this we see that $1-z$ is small in the entire region outside the throat. This limit (low temperature, low frequency and \eqref{AppBlimz}) makes both the frequency and angular momentum dependent terms in \eqref{eq:3dwave} negligible, and simplifies the solution to a simple power law proportional to $(1-z)^{\frac{\Delta_\pm}{2}}$. We can write these as 
\begin{equation}
	\chi \sim A \left(\frac{r^2-r_+^2}{\ell_3^2}\right)^{-\frac{\Delta_-}{2}} + B \left(\frac{r^2-r_+^2}{\ell_3^2}\right)^{-\frac{\Delta_+}{2}},
\end{equation}
 with coefficients $A,B$ depending on $\omega$ and $\ell$; because we reduce to the simple power law solutions, this is the only dependence on those variables. For $r\to\infty$, these coefficients match those of the asymptotic expansion \eqref{eq:AdS3expansion} near the AdS$_3$ boundary.
 
 To match to the AdS$_2$ boundary, we now need only expand this solution for small $r-r_+$, and match coefficients of powers of the variable $\frac{4}{\ell_3}(r-r_+)$ in the expansion \eqref{eq:AdS2expansion}:
\begin{equation}
	\tilde{A} = \left(\frac{r_+}{2\ell_3}\right)^{-\frac{\Delta_-}{2}} A,\quad \tilde{B} =  \left(\frac{r_+}{2\ell_3}\right)^{-\frac{\Delta_+}{2}} B
\end{equation}
 Writing this in terms of the right-moving temperature $\beta_R\sim \frac{\pi \ell_3}{r_+}$, and using $\Delta_+=\Delta$, $\Delta_-=2-\Delta$, we finally have
\begin{equation}
	\tilde{A} = \left(\frac{2}{\pi}\beta_R\right)^{1-\frac{\Delta}{2}} A,\quad \tilde{B} = \left(\frac{2}{\pi}\beta_R\right)^{\frac{\Delta}{2}} B
\end{equation}
 as stated in \eqref{eq:boun}. This relation is independent of where we put the cut-off between the 2d and 3d boundaries as expected.

\mciteSetMidEndSepPunct{}{\ifmciteBstWouldAddEndPunct.\else\fi}{\relax}
\bibliographystyle{utphys}
{\small \bibliography{references.bib}{}}

\end{document}